\documentclass[12pt,amsmath,amssymb,aps,nofootinbib]{revtex4-1}
\usepackage{graphicx,epsfig,dcolumn,bm,epic,eepic,float}% Include figure files
\usepackage{amsmath}
\usepackage[T1]{fontenc}
\usepackage[utf8]{inputenc}
\usepackage{mathtools} 
\usepackage{pifont}
\usepackage{amsmath}
\usepackage{latexsym}
\usepackage{color}
\usepackage{cancel}
\usepackage{scrextend}
\usepackage{amsmath}
\usepackage{amsthm}
\usepackage{eucal}
\usepackage{amssymb}
\usepackage{mathrsfs}
\usepackage[bottom]{footmisc}
\usepackage{makeidx,shortvrb,latexsym}
\usepackage{epstopdf}
\usepackage{mathtools}
\usepackage{ragged2e}
\usepackage{autobreak}
\usepackage[T1]{fontenc}
\usepackage{inputenc}
\usepackage{lmodern}

\newcommand{\newc}{\newcommand}
\newc{\gev}{\,GeV}
\newcolumntype{M}[1]{>{\centering\arraybackslash}m{#1}}
\newcolumntype{N}{@{}m{0pt}@{}}
\newc{\mev}{\,MeV}
\newc{\rpv}{$\mathrm{\not\!R_p}$}
\newc{\rp}{$\mathrm{R_p}$}
\newc{\real}{\mathcal{R}e}
\newc{\alsm}{{\displaystyle \sum_{\alpha=1,2}}}
\newc{\besm}{{\displaystyle \sum_{\beta=1,2}}}
\newc{\al}{\alpha}
\newc{\sgn}{\mr{sgn}\,}
\newc{\be}{\beta}
\newc{\ga}{\gamma}
\newc{\de}{\delta}
\newc{\sla}{\!\!\!\!\!\not\:\:\!}
\newc{\slab}{\!\!\!\!\!\not\,\,\,}
\newc{\slac}{\!\!\!\!\!\!\!\not\,\,\,\,}
\newc{\met}{$\not\!\!E_T$}
\newc{\cw}{\cos\theta_W}
\newc{\sw}{\sin\theta_W}
\newc{\ssw}{\sin^2\theta_W}
\newc{\ccw}{\cos^2\theta_W}
\newc{\cbe}{\cos\beta}
\newc{\sbe}{\sin\beta}
\newc{\ort}{\frac1{\sqrt{2}}}
\newc{\sh}{\hat{s}}
\newc{\uh}{\hat{u}}
\newc{\tha}{\hat{t}}
\newc{\sa}{\sin\al}
\newc{\ca}{\cos\al}
\newc{\mz}{M_{\mr{Z}}}
\newc{\mw}{M_{\mr{W}}}
\newc{\bv}{$\mathrm{\not\!B}$}
\newc{\lv}{$\mathrm{\not\!L}$}
\newc{\beq}{\begin{equation}}
\newc{\eeq}{\end{equation}}
\newc{\ie}{{\it i.e.\/}\ }
\newc{\lam}{\lambda}
\newc{\cht}{\tilde{\chi}}
\newc{\glt}{\tilde{g}}
\newc{\upt}{\tilde{u}}
\newc{\qkt}{\tilde{q}}
\newc{\elt}{\tilde{\ell}}
\newc{\hgt}{\tilde{H}}
\newc{\nut}{\tilde{\nu}}
\newc{\dnt}{\tilde{d}}
\newc{\ftl}{\mr{\tilde{f}}}
\newc{\psb}{\bar{\psi}}
\newc{\rtt}{2^{1/2}}
\newc{\mut}{\tilde{\mu}}
\newc{\mr}{\mathrm}
\newc{\bath}{\bar{\theta}}
\newc{\tht}{\theta}
\newc{\JC}{{\bf J}}
\newc{\lra}{\longrightarrow}
\newc{\eg}{{\it e.g.\  }}
\newc{\barr}{\begin{eqnarray}}
\newc{\earr}{\end{eqnarray}}
\newc{\me}{\mathcal{M}}
\newc{\dbm}{\partial_\mu}
\newc{\dbmu}{\stackrel{\leftrightarrow\  }{\partial^\mu}}
\newc{\sgm}{\sigma_\mu}
\newc{\captionB}[2]{\caption[{#1}]{{\small {#2}}}}
\pdfoutput=1

\numberwithin{equation}{section}
\begin{document}
\noindent{\hfill \small IPPP/21/59 \\[0.1in]}

\title{{\Large Electroweak Radiative Corrections in Precision LHC Measurements of $W^\pm/Z^0$+jets}\vspace*{5mm}}

\author{\sc Neda Darvishi$^{\,a,b}$} 
\author{\sc M.R. Masouminia$^{\,c}$\vspace*{3mm}}
\affiliation{$^a$Department of Physics and Astronomy, Michigan State University, East Lansing, MI 48824, USA}
\affiliation{$^b$Institute of Theoretical Physics, Chinese Academy of Sciences, Beijing 100190, China}
\affiliation{$^c$Institute for Particle Physics Phenomenology, Durham University, Durham DH1 3LE, United Kingdom}
    
\begin{abstract}
${}$

\centerline{\bf ABSTRACT} \medskip
\noindent

We calculate the fiducial and differential $W^{\pm}/Z^0+jet(s)$ production cross-sections in the presence of electroweak (EW) corrections through virtual loop contributions to the matrix elements (MEs) of the processes and real partonic cascade emissions. The calculations are carried out for proton-proton collisions at $\sqrt{s} = 13$~TeV, using \textsf{Herwig~7} general-purpose Monte-Carlo event generator with leading-order or next-to-leading-order MEs that are interfaced with different parton-shower configurations. The results are compared with precision experimental measurements from ATLAS collaboration and with similar predictions within the $k_t$-factorisation framework, providing a test for the validity of the newly-implemented \texttt{QCD$\oplus$QED$\oplus$EW} parton shower in \textsf{Herwig~7}. It is shown that the inclusion of EW radiations in the parton shower simulations improves \textsf{Herwig~7}'s predictions in describing the experimental data. Additionally, the inclusion of parton shower-induced real EW emissions can take precedence over the incorporation of virtual EW corrections for the simulation of EW-sensitive events.

\end{abstract}
\maketitle

\section{Introduction}
\label{sec:intro}

%importyance of the observables and their sensitivity to EW corrections
In the current era of precision high-energy particle physics, the processes involving electroweak (EW) gauge boson productions are amongst the most important tests of the Standard Model (SM) and perturbative QCD as they are sensitive to the properties of EW bosonic self-interactions. Henceforth, despite the fact that the QCD effects are typically dominant at the LHC events, there is an ever-increasing necessity to account for the sub-leading EW effects in particular regions of the phase-space, where these contributions become significant. Such observations would also serve as screening studies for the large backgrounds in the measurement of Higgs boson production and in searches for physics beyond the SM. Furthermore, the existence of large hadronic backgrounds substantially reduces the accuracy of the experimental measurements and therefore elevates the importance of performing precision calculations that simultaneously incorporate real and virtual EW corrections.

%importance of EW corrections
However, when aiming for full particle-level simulations as accomplished by general-purpose Monte-Carlo event generators, the consistent inclusion of EW corrections to the hard-scattering processes poses a severe theoretical challenge~\cite{Denner:2019vbn}, in particular in the context of multijet-merged calculations~\cite{Bellm:2017ktr,Gutschow:2018tuk,Brauer:2020kfv,Bothmann:2021led}. In \textsf{Herwig~7}~\cite{Bahr:2008pv,Bellm:2015jjp,Bellm:2017bvx}, the inclusion of the virtual contributions into the matrix elements (MEs) of the involving channels can be done using a number of conventional ME provides, e.g. \textsf{GoSam}~\cite{Cullen:2014yla}, \textsf{MadGraph}~\cite{Alwall:2014hca} with \textsf{MadLoops}~\cite{Hirschi:2011pa,Frederix:2018nkq} and \textsf{OpenLoops}~\cite{Buccioni:2019sur}. These virtual corrections are usually large and have negative signs. This is since the existence of incomplete infrared cancellations due to the broken structure of the gauge group introduces logarithms of the scale of the process and that of the EW scale, appending EW Sudakov logarithms~\cite{Beenakker:2000kb} which are negative and grow with the size of the kinematic invariants~\cite{Schonherr:2017qcj}. Considering these negative contributions in the theoretical predictions for the LHC have already resulted in up to $\mathcal{O}(\%20)$ corrections for the TeV-scale observables~\cite{Schonherr:2017qcj}. On the other hand, the integration of the QCD and EW shower algorithms links real EW radiation to the negative EW virtual corrections by introducing $W^\pm/Z^0$ boson production sequences inside QCD jets~\cite{Christiansen:2014kba}, allowing one to treat the real and virtual radiations of the EW bosons on equal footing. Moreover, it is expected that the SM heavy particles like top-quarks and Higgs bosons will behave as massless partons as the probe energy of the collider grows much larger than their masses~\cite{CMS:2019kqw,Aaboud:2018gay}, which in turn promotes the need for using parton shower algorithms capable of performing process-independent EW enhancements in addition to the conventional QCD and QED emissions. To this end, some attempts have been made for adding EW radiations to the existing parton shower algorithms~\cite{Chiesa:2013yma,Christiansen:2014kba,Krauss:2014yaa,Chen:2016wkt,Mangano:2002ea,Kleiss:2020rcg,Brooks:2021kji,Pagani:2021vyk}. Nevertheless, \textsf{Herwig~7} is so far the only conventional general-purpose event generators that employs a complete and process-independent EW parton shower to realize a \texttt{QCD$\oplus$QED$\oplus$EW} level enhancement and treat the full scope of high-energy collinear EW physics~\cite{Masouminia:2021kne}.

%introduction of EWPS in Heriwg
In~\cite{Masouminia:2021kne}, the necessary steps for the implementation of an angularly-ordered (AO) initial- and final-state EW parton shower in \textsf{Herwig~7} have been discussed in length, including the introduction of all relevant \textit{quasi}-collinear helicity-dependent splitting functions. The resulting \texttt{QCD$\oplus$QED$\oplus$EW} parton shower algorithm covers the full range of final-state EW emissions while capturing the most relevant parts of the initial-state corrections, avoiding the use of numerically expensive and physically insignificant initial EW self-interactions\footnote{The inclusion of the initial-state EW radiations, where the EW gauge bosons can be considered as intermediate steps in the evolution of the incoming partons, would require the use of EW PDF in the backward convolution of the beam particles and is not currently implemented in \textsf{Herwig~7} due to being numerically expansive while not being stable and reliable~\cite{Masouminia:2021kne}.}. While this novel algorithm has undergone a plethora of validation checks in both performance and physics, it would be very interesting to further investigate the effects of employing the full range of \textsf{Herwig~7}'s \texttt{QCD$\oplus$QED$\oplus$EW} parton shower in comparison with the conventional \texttt{QCD$\oplus$QED} cascades, with or without the presence of virtual EW corrections in the MEs of some EW-sensitive mile-stone processes to examine its effectiveness in predicting the recent precision measurements at the LHC~\cite{ATLAS:2020juj,ATLAS:2021jgw}. 

%Herwig in collinear fact vs kt-fact and its importance
Henceforth, in this work, we first calculate the fiducial and differential cross-sections for the production of $W^+W^-$ pairs in association with one or more hadronic jets within the collinear factorization framework. The simulations are performed using the leading-order (LO) and next-to-leading-order (NLO) MEs in addition to the corresponding virtual EW corrections up to one loop and showered with either \texttt{QCD$\oplus$QED} or \texttt{QCD$\oplus$QED$\oplus$EW} type parton showers in \textsf{Herwig~7}. Secondly, we look at the production rate of a single $Z^0$ boson accompanied by one or more $b$-tagged jets. In this case, our NLO predictions are generated with or without the inclusion of real and virtual EW corrections, allowing us to investigate different settings in the evaluation of EW-sensitive events in \textsf{Herwig~7}. The predicted signals are compared with the experimental measurements from the ATLAS collaboration~\cite{ATLAS:2020juj,ATLAS:2021jgw} and against third-party theoretical predictions within the $k_t$-factorisation framework~\cite{Catani:1990xk,Catani:1990eg,Catani:1993ww}, using the unintegrated parton distribution functions (UPDFs) of Kimber-Martin-Ryskin (KMR)~\cite{Kimber:2001sc,Martin:2009ii}. The latter framework has been used as a control signal since its validity in accurately predicting the behaviour of experimental observations has been repeatedly proven to be independent of the use parton shower enhancements~\cite{Modarres:2016hpe,Modarres:2016tow,Darvishi:2016fwo,Modarres:2018dwj,Darvishi:2019uzp,Darvishi:2020paz}, while being computationally expensive and unsuitable for use in general-purpose event generators~\cite{Darvishi:2022gqt,Campbell:2022qmc,Frixione:2022ofv,Feng:2022inv}. Note that the production of $W^\pm$ and $Z^0$ bosons in association with hadronic jets have been studied extensively in the literature, e.g. in Refs.~\cite{Brauer:2020kfv,Gauld:2020deh}.

%CASCADE3
At this point, we should note that a more balanced and homogeneous comparison between the collinear and $k_t$-factorization effects could be obtained by comparing \textsf{Herwig~7} results to predictions from \textsf{CASCADE3} Monte-Carlo event generator~\cite{Baranov:2021uol,Jung:2010si}. This is because these are both parton shower generators that take into account angular ordering effects via colour-coherence radiations. \textsf{CASCADE3} has been used recently in the analysis of $Z^0+jet(s)$ productions in \cite{BermudezMartinez:2022bpj,Yang:2022qgk}, while early comparisons with \textsf{Herwig~7} were performed for the case of jet production in \cite{Hautmann:2008vd}.
However, in the context of this work and for the sake of simplicity, we instead adopted the KMR \texttt{UPDF} method to produce our $k_t$-factorization predictions.

%layout of the paper
This paper is organised as follows: In Section \ref{Fram}, we briefly discuss our computational framework in both collinear and $k_t$-factorisation schemes and discuss different contributions in $p p \to W^+ W^- + \;jet(s)$ and $p p \to Z^0 + \;jet(s)$ events at the LHC. This section also includes a description of the AO EW parton shower in \textsf{Herwig~7} as well as the KMR \texttt{UPDF} and their properties. In Section \ref{Num}, we review the numerical methods and computational choices while our results are presented in Section \ref{Res}. Finally, Section \ref{Conc} is devoted to our summary and conclusions.

\section{The Framework}\label{Fram}

\begin{figure}[t]
\centering
\includegraphics[width=1\textwidth]{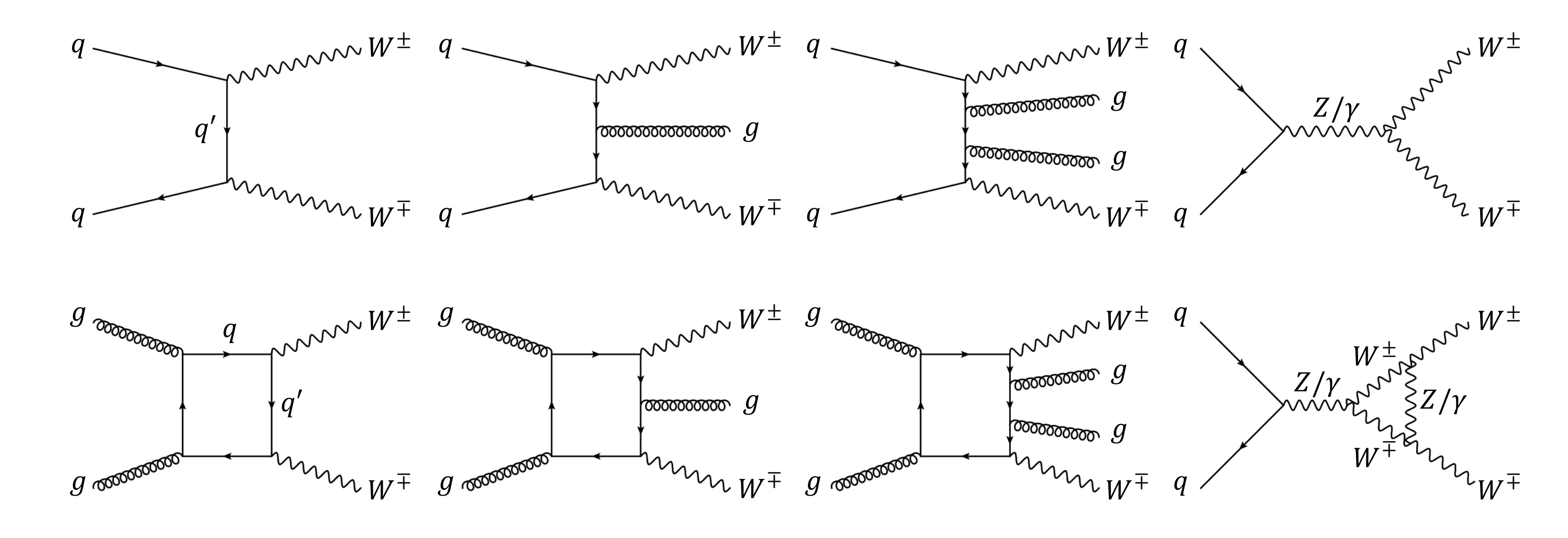}
\caption{\it Born-level (top row) and one-loop (bottom row) contributions to the production of $W^+W^-$ boson pairs alongside hadronic jets. The last channels in each row represent pure EW contributions to the MEs.}
\label{WW-diags}
\end{figure}

%%%%%%%%%% WW %%%%%%%%%%
In the current LHC energies, the leading processes that contribute to the production of $W^+W^-$ pairs are the Born-level $q\bar{q} \to W^+W^-+jet(s)$ and the loop-induced gluon–gluon fusion processes, $gg \to W^+W^-+jet(s)$. Additionally, the $q\bar{q} \to W^+W^-$ processes can also contribute in $W^+W^-+jet(s)$ production through parton shower-induced initial- and final-state radiations. Representative diagrams for these production channels are shown in Fig.~\ref{WW-diags}. The full list of relevant processes also includes non-dominant pure EW contributions in both LO and NLO levels, represented on the right-hand side of each row in Fig.~\ref{WW-diags}. The processes that involve the exchange of Higgs bosons would also appear in such events, mainly through $gg \to H \to W^+W^-$ resonants but these can be strongly suppressed through kinematic constraints. In contrast, EW virtual contributions through gauge boson self-couplings (as shown in Fig.~\ref{WW-diags}) cannot be kinematically suppressed and would also affect the calculated predictions through destructive interferences in the MEs of $pp \to W^+W^-+jet(s)$. 

\begin{figure}[t]
\centering
\includegraphics[width=1\textwidth]{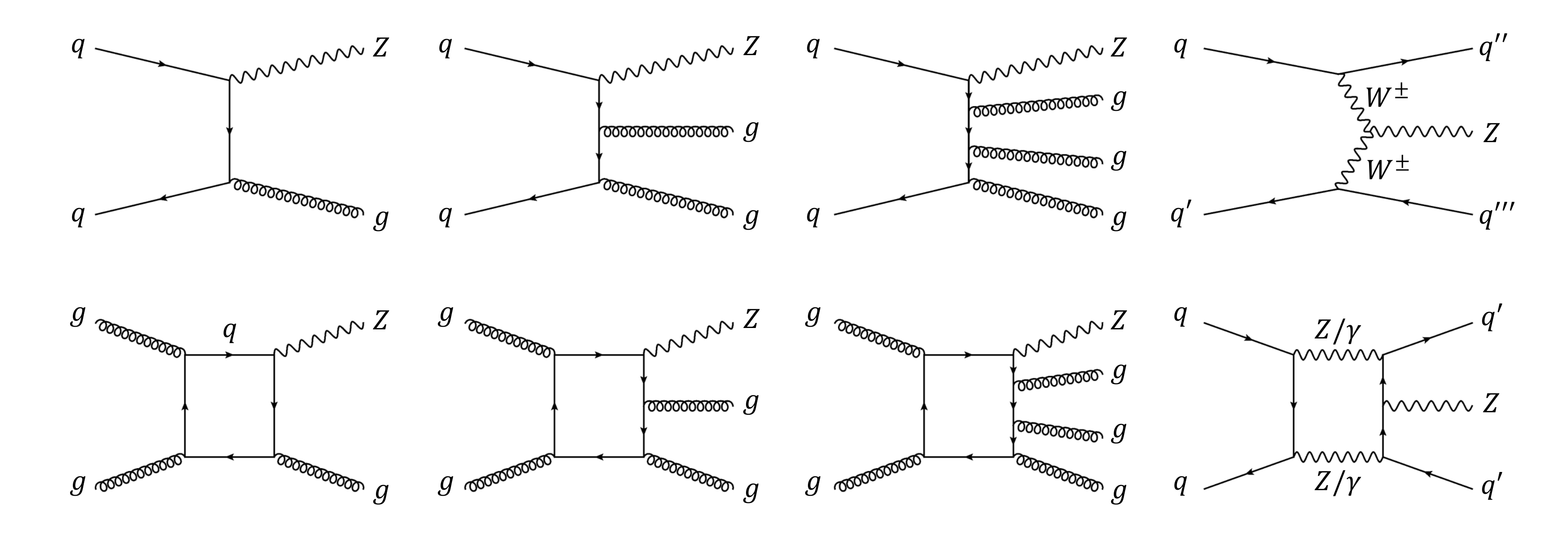}
\caption{\it Born-level (top row) and one-loop (bottom row) contributions to the production of $Z^0$ bosons alongside hadronic jets. The last channels in each row represent pure EW contributions to the MEs.}
\label{Z-diags}
\end{figure}

%%%%%%%%%% Z %%%%%%%%%%
Similar to the production of $W^+W^-$ pairs, the production of $Z^0$ bosons alongside $b$-tagged hadronic jets is sensitive to real and virtual EW corrections and can serve as an important test for the treatment of these effects~\cite{Gauld:2020deh}. In addition, the $Z^0 + jet(s)$ events may be augmented by EW production of $b$-tagged dijets, a class of processes that are extremely sensitive to the weak vector-boson scattering production mechanism and provides a stringent test of the SM gauge structure~\cite{ATLAS:2020juj}. Fig.~\ref{Z-diags} depicts the representative diagrams for the LO and NLO productions of $Z^0+jet(s)$. Again, the QCD contributions are dominant and pure EW diagrams are expected to be sub-leading. Moreover, working within the 5-flavour scheme would allow additional contributions from EW diagrams and from the initial-state jets through the parton distribution functions (PDFs) of the protons. 

Given the sensitivity of the $W^+W^-+jet(s)$ and $Z^0+jet(s)$ production events, they can provide a benchmark to test the predictions of Monte-Carlo simulations, especially concerning their treatment of EW real emissions and virtual ME corrections. Previously, such tests would have been limited to only the virtual corrections in the MEs of the relevant sub-processes, and no viable and process-independent EW parton shower was available. Now, with the recent implementation of the AO EW parton shower in \textsf{Herwig~7}~\cite{Masouminia:2021kne}, it would be very interesting to compare the effects of both real and virtual corrections in uniform simulations. 

The above-mentioned processes are also inherently sensitive to the choice of PDFs and can be even used to constrain them. This would create an intrinsic ambiguity in the calculations that may be reduced with the inclusion of higher-order perturbative QCD and EW corrections. Nevertheless, it would be instructive to produce an equivalent set of predictions using the UPDFs of KMR that are less sensitive to the variations of PDFs~\cite{Modarres:2016hpe}. In such a framework, UPDFs are used instead of PDFs to weight the relevant ladder-type partonic subprocesses, while the MEs are calculated using the eikonal approximation for the incoming quark spin-densities and the non-sense polarization approximation for the polarization vectors of the incoming gluons~\cite{Modarres:2016hpe,Baranov:2008hj}, in addition to suppressing soft gluon singularities by employing an AO constraint.

One should note that the inclusion of the one-loop QCD channels is important for the EW gauge boson production events in the collinear factorization framework, particularly since they cause large destructive interferences. However, in the $k_t$-factorization formalism, summing tree-level and QCD loop-induced processes will cause irreducible double-counting since the box and crossed-box diagrams are used in the definitions of the UPDFs. Therefore, one can use either the Born-level or the one-loop QCD channels in association with the UPDFs of $k_t$-factorization, and not both.

In the following subsections, we briefly outline the details of the EW parton shower in \textsf{Herwig~7} and the framework of $k_t$-factorisation.

\subsection{Angularly-Ordered Electroweak Parton Shower In \textsf{Herwig~7}}\label{HwSet}

The implementation of an AO initial-state (IS) and final-state (FS) EW parton shower in \textsf{Herwig~7} has been done via the insertion of quark splittings, $q\to q'V$ and $q\to qH$ as well as the EW gauge boson splittings, $V \to V'V''$ and $V \to V H$ in \textsf{Herwig~7}'s AO parton shower algorithm~\cite{Masouminia:2021kne}. These, alongside the previously existing $H \to q\bar{q}$, $H \to VV^*$ and $V \to q \bar{q}'$ decay modes, have created a satisfactory picture of the IS and FS EW radiations in the simulated events. The {\it quasi}-collinear spin-averaged splitting function for a generic splitting, $\widetilde{ij} \to i+j$, was calculated by summing over the helicity states $\lambda$ as
\begin{equation}
P_{\widetilde{ij}\to i+j} (z,\tilde{q}) = 
	\sum_{\lambda} \left| H_{\widetilde{ij}\to i+j}(z,\tilde{q};\lam) \right|^2,
\end{equation}
with $z$ and $\tilde{q}$ as the light-cone momentum fraction and the evolution scale. Here, the helicity amplitudes, $H_{\widetilde{ij}\to i+j}(z,\tilde{q};\lam)$ may be defined using the Feynman rules for the given splitting vertex~\cite{Masouminia:2021kne}. The resulting splitting functions can be used to calculate the differential cross-section of the production of the children particles $i$ and $j$ in the \textit{quasi}-collinear limit as 
\begin{equation}
d \sigma_{i+j} \simeq 
{\alpha_{\rm int}(\tilde{q}^2) \over 2 \pi} {d\tilde{q}^2 \over \tilde{q}^2} dz \;
P_{\widetilde{ij} \to i + j}(z,\tilde{q}) \; d \sigma_{\widetilde{ij}},
\label{collaprox}
\end{equation}
with $\alpha_{\rm int}$ as the relevant running coupling constant. The express analytic forms of utilized EW splitting functions are presented in~\cite{Masouminia:2021kne} and the resulting \texttt{QCD$\oplus$QED$\oplus$EW} parton shower has successfully undergone performance tests and physics validity checks. 

\subsection{$\bf k_t$-Factorisation Framework and KMR UPDFs}\label{UPDF}

In the $k_t$-factorization framework, the differential cross-section for the production of $W^+W^-+jet(s)$ and $Z^0+jet(s)$ events can be written as
\begin{widetext}
\begin{eqnarray}
    d\sigma &=& 
    \sum_{a,b=q,g} \int {dx_1 \over x_1} {dx_2 \over x_2}
    {dk_{1,t}^2 \over k_{1,t}^2} {dk_{2,t}^2 \over k_{2,t}^2}
    \; f_{a}(x_1,k_{1,t}^2,\mu^2)\; f_{b}(x_2,k_{2,t}^2,\mu^2) \; 
    \nonumber \\ &\times&
    {d\varphi \big( ab \to leptons+jet(s) \big) \over x_1 x_2 s}
	\big| {\mathcal{M}} \big( ab \to leptons+jet(s) \big)\big|^2 ,
	\nonumber \\
    \label{TCS}
\end{eqnarray}
\end{widetext}
with the particle phase-space 
	\begin{eqnarray}	
	d\varphi &=& (2\pi)^4
	\prod_{i\in \text{final-state}} \left[ {1 \over 16\pi^2} \; dp_{i,t}^2 \; dy_i \; {d\phi_i \over 2\pi} \right]
	%\nonumber \\ & \times &	
	\delta^{(4)} \left( k_1 + k_2 -\sum_{j\in \text{final-state}} p_j \right).
    \label{dPHI}
	\end{eqnarray}
In the above equations, $x_{i}$, $k_{i,t}$ and $\mu$ are the longitudinal momentum fraction and the transverse momenta of the incoming partons and the factorisation scale of the sub-process, respectively, while $\sqrt{s}$ is the centre-of-mass energy. The MEs are denoted by $\mathcal{M}$ while $y_i$ and $\phi_i$ are the pseudorapidities and the angles of emission of the particles. Also, $k_i$ and $p_i$ are the momenta of the incoming and produced particles. Here, $x_{1}$ and $x_{2}$ may be given as
\begin{eqnarray}
x_{1,2} = {1 \over \sqrt{s}} \sum_{i\in \text{final-state}}  m_{i,t} e^{\pm y_i},
\end{eqnarray}
with the the transverse mass of the final state particles, $m_{i,t} = (m_i^2 + p_{i,t}^2)^{1/2}$. 

The KMR UPDFs are defined as \cite{Kimber:2001sc}
\begin{eqnarray}
f_a(x,k_t^2,\mu^2) = \Delta_S^a(k_t^2,\mu^2) 
\sum_{b=q,g} \left[ {\alpha_S(\mu^2) \over 2\pi}
\int^{\Delta_{\rm AO}}_{x} dz \; P_{ab}(z) \; b\left( {x \over z}, k_t^2 \right) \right] , 
\label{KMR_UPDF}
\end{eqnarray}
using the Sudakov form factor, 
\begin{eqnarray}
\Delta_S^a(k_t^2,\mu^2) =
  exp \left( - \int_{k_t^2}^{\mu^2} {\alpha_S(k^2) \over 2\pi}
    {dk^{2} \over k^2} \sum_{b=q,g} \int^{\Delta_{\rm AO}}_{0} dz' P_{ab}(z') \right),
   \label{SFF}
\end{eqnarray}
with $P_{ab}(z)$ as the LO splitting functions for $b \to a+X$ partonic splittings \cite{Kimber:2001sc} and $\Delta_{\rm AO} = \mu / (\mu+k_t)$ as the AO cut-off hat defines the kinematics of the KMR UPDFs. To numerically calculate these UPDFs, one needs to use the conventional PDFs, denoted by $b( x, k_t^2)$. In this study, these PDFs are obtained from the \texttt{CT14} PDF libraries~\cite{Dulat:2015mca,Hou:2016sho}.

The MEs of the relevant sub-processes are calculated using the Feynman rules in combination with the eikonal and the non-sense polarization approximations for the incoming quarks and gluons~\cite{Modarres:2018dwj,Baranov:2008hj}. One should note that these MEs need to be calculated up to the leptonic levels in order to produce comparable results with their counterparts in \textsf{Herwig~7}. This, in addition to the complexity of numerical calculation of the $k_t$-factorisation UPDFs, makes them unsuitable for general-purpose Monte-Carlo event generators. 

%comment on the limitations of the $k_t$-factorisation framework
At this point, it would be appropriate to comment on the limitations of the $k_t$-factorisation. One major issue in this framework is the scale-dependency of its UPDFs through the AO cut-off, $\Delta_{\rm AO}$, in the integrated PDFs and the Sudakov form factors, Eqs. (\ref{KMR_UPDF}-\ref{SFF}). It has been shown that this scale-dependent behaviour, which is much more pronounced in this framework compared to the collinear factorisation, could drastically change the resulted predictions, see e.g.~\cite{Modarres:2016phz}. Meanwhile, employing a strong-ordering cut-off $\Delta_{\rm SO} = k_t / \mu$, leads to large discrepancies compared to the AO choice~\cite{Golec-Biernat:2018hqo}. Another concern, particular to the KMR UPDFs, is the simultaneous use of the AO cut-off for both gluon and quark successive emissions. This is in fact miss-aligned with the colour-coherent radiation effects that are meaningful only for successive real gluon radiations~\cite{Martin:2009ii}. Opting out quark terms from the AO constraint results in alternative UPDFs that are less predictive of the data~\cite{Modarres:2016hpe,Modarres:2016phz}. On the other hand, the KMR UPDFs are spawned by convoluting the conventional PDFs by single final-step emissions. This could enforce limitations on the phase-space of these UPDFs, resulting in significant discrepancies with their more recent counterparts~\cite{Hautmann:2019biw,Guiot:2019vsm}, e.g. from the Parton Branching\footnote{Parton Branching and a number of other transverse momentum dependent parton distributions, which can be used as KMR alternatives, are available in the \textsf{TMDlib2} and \textsf{TMDplotter} libraries~\cite{Hautmann:2014kza,Abdulov:2021ivr}.} formalism~\cite{Hautmann:2017xtx,Hautmann:2017fcj,BermudezMartinez:2019anj,Martinez:2021chk}. Regardless of the above-mentioned limitations, it has been shown that the KMR framework as an effective theory is capable of predicting a wide variety of experimental measurements, within appropriately chosen kinematic settings. 

\section{Numerical Analysis}\label{Num}

%%%%%%%%%%%%%%%%% W %%%%%%%%%%%%%%%%%
In this paper we aim to analyse $W^+W^-+jet(s)$ and $Z^0+jet(s)$ productions in proton-proton collisions at $\sqrt{s} = 13$~TeV centre-of-mass energy based on the recent measurements of the ATLAS collaboration~\cite{ATLAS:2020juj,ATLAS:2021jgw}. The Monte-Carlo simulations are carried out in the 5-flavour scheme, using two different frameworks:

(i) \textbf{The collinear factorisation framework}, using \textsf{Herwig~7}. Here, the LO and NLO MEs are calculated with and without virtual EW contributions, using \textsf{MadGraph5} (v3.1.0), up to one QCD and/or EW loop. For the case of $W^+W^-+jet(s)$ production, both LO and NLO MEs are generated up to two explicit jets and include EW virtual corrections with the same accuracy levels. 
Multi-jet merging has been done using \textsf{Herwig~7}'s \textsf{FxFx} interface~\cite{Bellm:2015jjp} that supports \texttt{MC@LO} and \texttt{MC@NLO} matchings through \texttt{MatchBox}~\cite{Platzer:2011bc} via Les~Houches-accord event (LHE) files generated by \textsf{MadGraph5}. 
Both cases have been showered with \texttt{QCD$\oplus$QED} or \texttt{QCD$\oplus$QED$\oplus$EW} parton showers, creating four different sets of results that are labelled accordingly. As an example, the results with LO MEs that are enhanced with a \texttt{QCD$\oplus$QED} parton shower are labelled as \texttt{Hw7.2 LO $\otimes$ QCD$\oplus$QED}. Comparing these results would allow us to investigate the effects of EW real emissions in a sensitive framework that also incorporates EW virtual contributions. 
In the $Z^0+jet(s)$ production case, MEs are generated up to one loop and 3 explicit jets, with and without NLO EW virtual corrections. The selection criteria require at least one $b$-tagged jet that can originate either explicitly or from the preceding partonic cascades. The results are again classified into four distinct sets, depending on the type of parton shower scheme. For example, the predictions that include both real and virtual EW corrections are labelled as \texttt{Hw7.2 S$\oplus$W $\otimes$ QCD$\oplus$QED$\oplus$EW} while the predictions in the absence of virtual EW contributions are marked as \texttt{Hw7.2 S $\otimes$ QCD$\oplus$QED$\oplus$EW}. 
The calculations are carried out using \textsf{Herwig-7.2.2}'s default particle data while setting the renormalisation and factorisation scales to be the sum of the transverse momenta of all final state particles. The resulting events are analysed by \textsf{Rivet} (v3.1.5) using \texttt{ATLAS\_2021\_I1852328} and \texttt{ATLAS\_2020\_I1788444} plugins~\cite{Buckley:2010ar}. 

Here we should note that since \textsf{MadGraph5} is currently incapable of producing unweighted events with NLO EW corrections, one has to extract the NLO QCD+EW signature indirectly, using pure NLO QCD samples in addition to the corrections obtained from $\sigma_{\rm QCD+EW} / \sigma_{\rm QCD}$. Such an enhancement has to be done bin-by-bin to produce NLO QCD+EW samples with the required accuracy. The resulting corrections can then be directly inserted in the analysis handler. This, however, could create ambiguity for the results in the absence of a well-defined matching between the NLO QCD+EW samples and the EW parton shower. Nevertheless, in the current LHC energy scales, the first few stages of the initial-state parton shower are dominated by QCD gluonic splittings. This would mean that the majority of EW splitting occurs as $V \to V'V''$ or as non-prompt $q \to q'V$ radiations and therefore, the double counting between the fixed-order and the showered samples is expected to be small. Indeed, when such a NLO EW matching becomes available, it would be interesting to study its effect on these energy scales. 

(ii) \textbf{The $k_t$-factorisation framework}, using KMR UPDFs that are introduced in Section~\ref{UPDF}. To carry out these computations, we have chosen the hard-scale the processes to be
\begin{equation}
    \mu^2 = \sum_{i \in \text{final-state}} p_{i,t}^2,
\label{HS}
\end{equation}
where $p_{i,t}$ are the transverse momenta of the fermionic final-state particles, i.e. quarks as jet progenitors and final-state leptons. Also, since the UPDFs will vanish at the limit $k_{t}^2\gg\mu^2$, we can set the bounds in the $k_{i,t}$ interactions of Eq.~\eqref{TCS} to be $[0 ,k_{t}^{\text{max}}]$ with
\begin{equation}
    k_{t}^{\text{max}} \equiv 4 \left[ \sum_{i \in \text{final-state}} p_{i,t}^{2,\text{max}} \right]^{1/2}.
    \label{ktmax}
\end{equation}
Note that for the non-perturbative domain of $k_{i,t} \in [0 , 1 \text{ GeV}]$, $f_{a}(x,k_{t}^2,\mu^2)$ takes on the form 
\begin{equation}
    f_a(x,k_{t}^2<\mu_0^2,\mu^2) = {k_{t}^2 \over \mu_0^2} a(x,\mu_0^2) \Delta_S^a(\mu_0^2,\mu^2), 
\end{equation}
and therefore $\lim_{k_{t}^2 \rightarrow 0} f_{a}(x,k_{t}^2,\mu^2) \sim k_{t}^2$. The relevant MEs for the $k_t$-factorisation simulations are generated up to five parton-level jets, using the algebraic manipulation toolkit \textsf{FORM}~\cite{FORM} and are checked independently by \textsf{MATHEMATICA}.

In both frameworks, we use the \texttt{CT14} PDF libraries~\cite{Dulat:2015mca,Hou:2016sho} and set the event selection constraints according to the specifications of the corresponding experimental measurements as expressed in~\cite{ATLAS:2020juj,ATLAS:2021jgw}.

\section{Results and Discussion}\label{Res}

In this section, we present the numerically calculated fiducial and differential cross-sections for $W^+W^-+jet(s)$ and $Z^0+jet(s)$ production events through $W^+ W^- \to e^+ \nu_e \; \mu^- \bar{\nu}_\mu$ and $Z^0 \to \ell^+ \ell^-/\nu_\ell \bar{\nu}_\ell$ decay modes at $\sqrt{s}= 13$~TeV, with the phase-space of the Monte-Carlo calculations defined in Table~\ref{tab1}. In all simulations, particle-level jets are identified by applying the anti-$k_t$ algorithm with $\Delta R(\ell, j) \geq 0.4$.

\begin{table}[t]
\centering
\begin{tabular}{| c | c c c c c c c c c c |}
\hline \hline
Event & & $p_\top^{\ell}$ & \hspace{0.3cm} &  $|y_{\ell \ell}|$ & \hspace{0.3cm} & $p_\top^{j}$ & \hspace{0.3cm} & $|y_{j}|$ & \hspace{0.3cm} & $N_{\rm jets}$ \\ \hline
%%%
$W^+W^-+jet(s)$ & & $>27$~GeV & & $<2.5$ & & $>30$~GeV & & $<4.5$ & & $\geq 1$ \\
%%%
$Z^0+jet(s)$    & & $>27$~GeV & & $<2.5$ & & $>20$~GeV & & $<2.5$ & & $\geq 1$ $b$-tagged
\\ \hline \hline
\end{tabular}
\caption{ \it Kinematic criteria defining the phase-space for the Monte-Carlo calculations of fiducial and differential cross-sections in $W^+W^-+jet(s)$ and $Z^0+jet(s)$ productions events.}
\label{tab1}
\end{table}

\begin{figure}[t]
\centering
\includegraphics[width=1\textwidth]{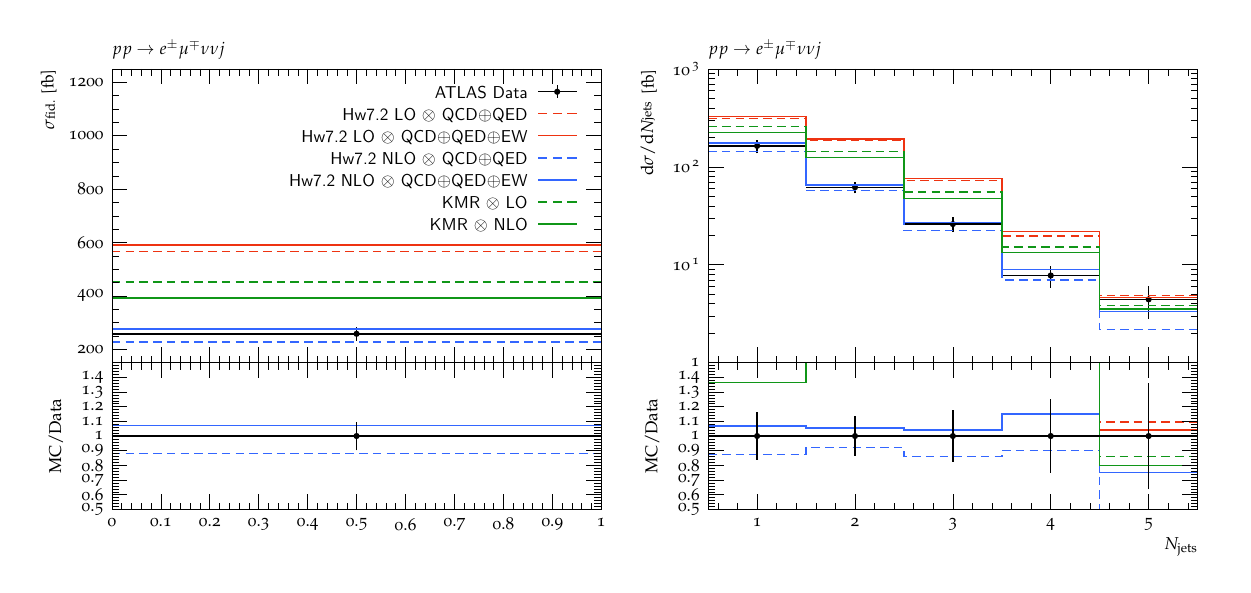}
\caption{\it Comparison of the fiducial and differential $W^+W^-+jet(s)$ cross-sections in various theoretical frameworks with ATLAS data~\cite{ATLAS:2021jgw}. Theoretical predictions from \textsf{Herwig~7} with LO and NLO MEs are respectively plotted with red and blue histograms and are showered with either the \texttt{QCD$\oplus$QED} (dashed lines) or the \texttt{QCD$\oplus$QED$\oplus$EW} (solid lines) parton shower schemes. The results in the $k_t$-factorisation, shown in green, are calculated using LO (dashed lines) or NLO (solid lines) MEs. All theoretical frameworks incorporate EW virtual corrections in the calculation of their MEs. The left panel presents a fiducial cross-section of $W^+W^-+jet(s)$ production while the right panel depicts the differential cross-section as a function of the number of produced jets.}
\label{W3}
\end{figure}

Fig.~\ref{W3} shows the fiducial and differential cross-sections for the production of $W^+W^-$ pairs alongside at least one hadronic jet, in both the collinear and the $k_t$-factorisation frameworks. The collinear predictions are computed using \textsf{Herwig~7} with either LO or NLO MEs, plotted in red and blue respectively. These results are enhanced via the implementation of an IS+FS parton shower in the \texttt{QCD$\oplus$QED} (dashed lines) or the \texttt{QCD$\oplus$QED$\oplus$EW} (solid lines) schemes. The $k_t$-factorisation predictions are plotted with green histograms by the use of either purely tree-level MEs (dashed lines marked as \texttt{KMR $\otimes$ LO}) or purely QCD and EW loop-induced MEs (solid lines marked as \texttt{KMR $\otimes$ NLO}). The left panel is a comparison between the fiducial cross-section of $W^+W^-+jet(s)$ production in various theoretical frameworks with ATLAS data~\cite{ATLAS:2021jgw}. The right panels show this comparison for the predicted differential cross-sections as functions of the number of produced hadronic jest, $N_{\rm jets}$. In both plots, ATLAS data favours NLO \textsf{Herwig~7} predictions that are enhanced with the use of \texttt{QCD$\oplus$QED$\oplus$EW} parton shower, although the NLO \textsf{Herwig~7} without real EW radiations also produce predictions within acceptable margins. 

Notably, in Fig.~\ref{W3}, the LO \textsf{Herwig~7} predictions overestimate the measurements by a factor of $\sim 2$. This can clearly showcase the importance of including loop-induced channels and the consequence of their destructive interference. The additional enhancement from the inclusion of EW radiations in the LO predictions is $\sim +2\%$ that increases to $\sim +10\%$ in the NLO case. This difference is expected to be more pronounced in the large-transverse-momentum tails, e.g. for $W^+W^-+ \geq 3 jets$, reaching $\sim +20\%$ for the NLO predictions with $W^+W^-+ 5 jets$. On the other hand, the $k_t$-factorisation predictions in both LO and NLO accuracies are better than the LO collinear results, overs estimating the ATLAS measurements by $\sim +30\%$ in the LO and $\sim +22\%$ in the NLO. 

\begin{figure}[t]
\centering
\includegraphics[width=1\textwidth]{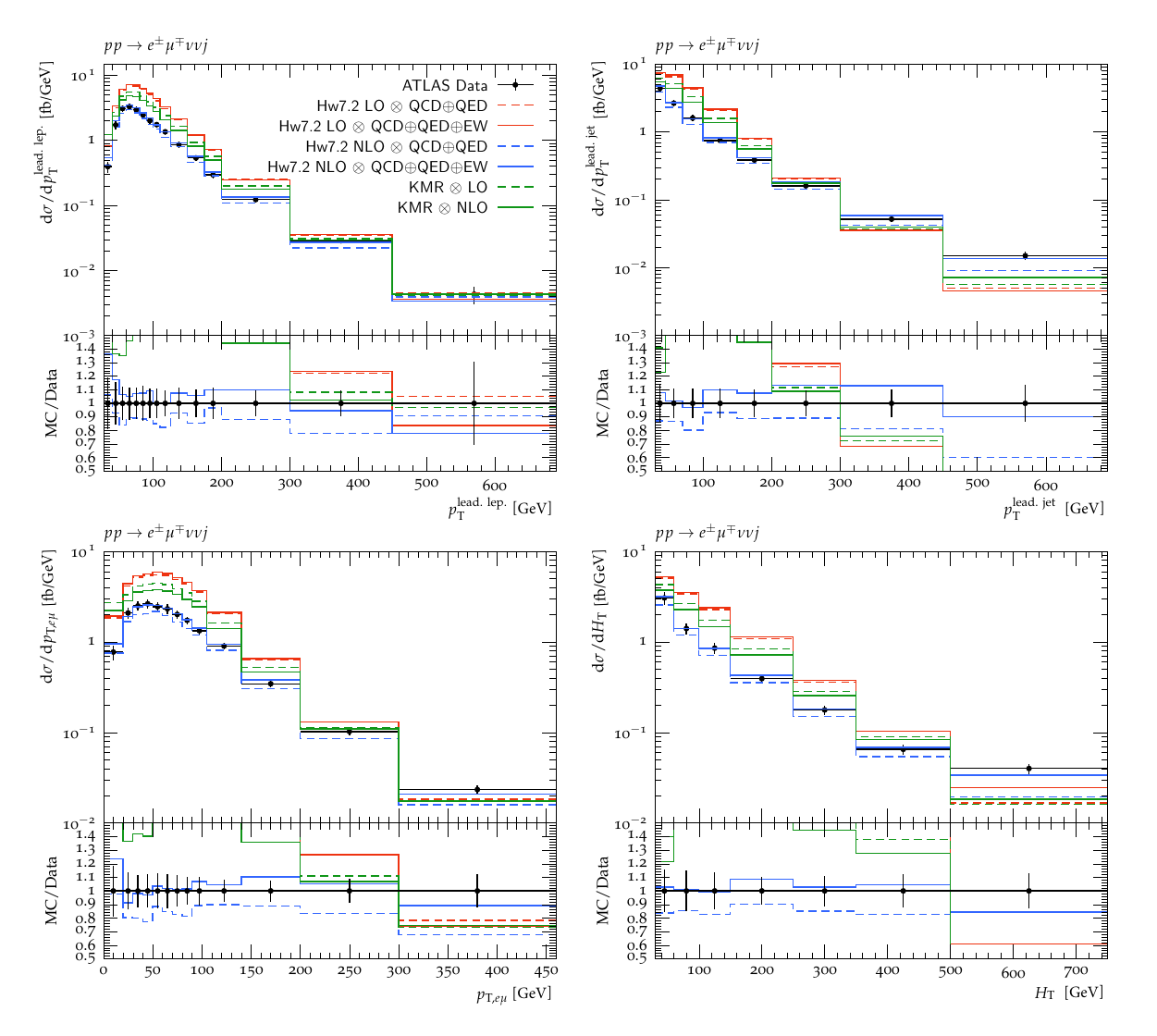}
\caption{\it Comparison of the differential $W^+W^-+jet(s)$ cross-section in various theoretical frameworks with ATLAS data~\cite{ATLAS:2021jgw}. The differential $W^+W^-+jet(s)$ cross-sections are plotted as functions of the transverse momenta of the leading lepton (top right), jet (top left) and the lepton pair $p_{\top, e\mu}$ (bottom left) and the scalar sum of all jet transverse momenta $H_\top$ (bottom right). The notation of the figure is the same as in Fig.~\ref{W3}.}
\label{W4}
\end{figure}

\begin{figure}[t]
\centering
\includegraphics[width=1\textwidth]{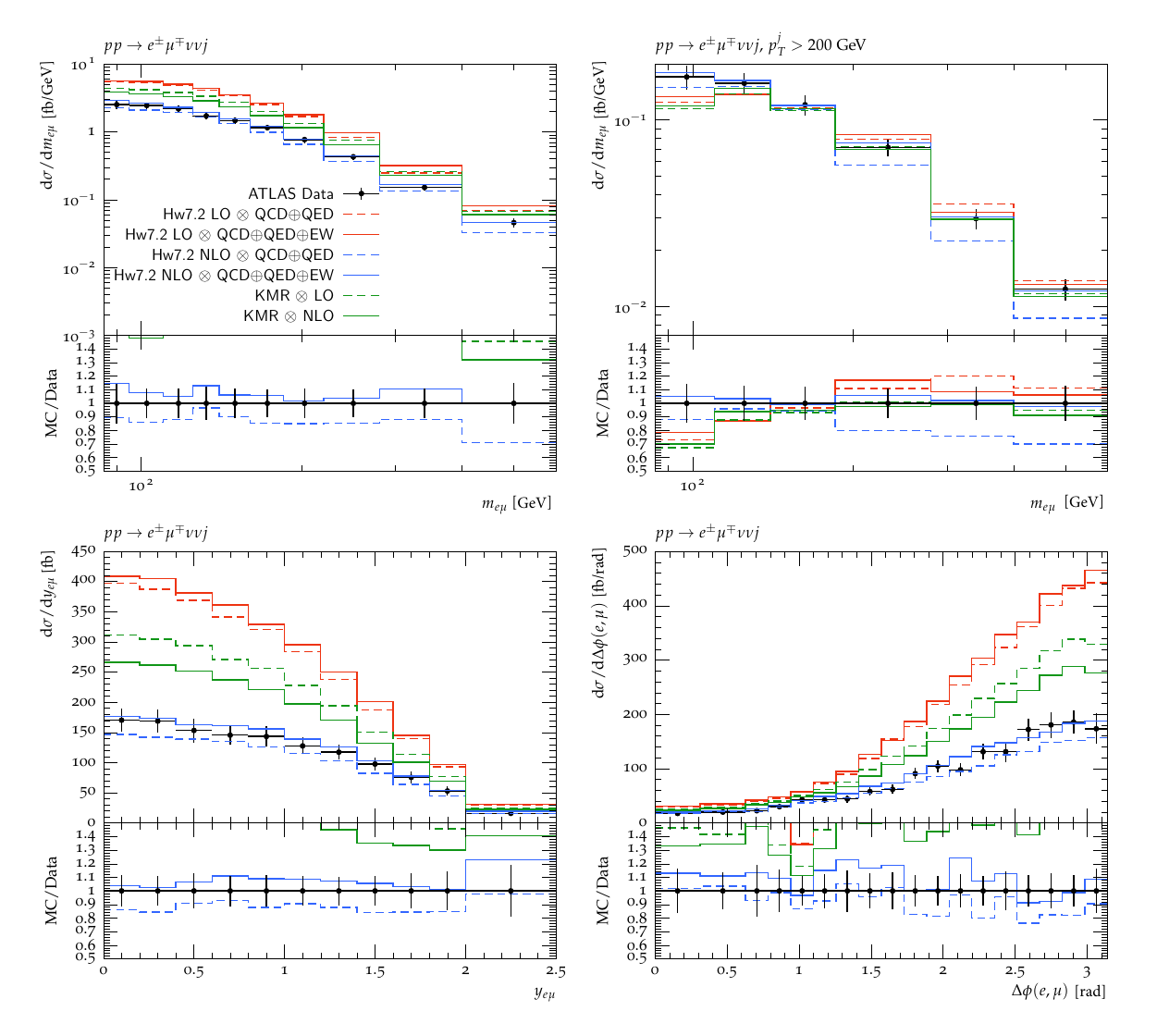}
\caption{\it Comparison of the differential $W^+W^-+jet(s)$ cross-section in various theoretical frameworks with ATLAS data~\cite{ATLAS:2021jgw}. The differential $W^+W^-+jet(s)$ cross-sections are plotted as functions of the invariant mass of the lepton-pairs for $p_\top^{j} > 30$~GeV (top right) and $p_\top^{j} > 200$~GeV (top left), the pesudorapidity of the lepton-pairs $p_{\top, e\mu}$ (bottom left) and the azimuthal separation of the two leptons $\Delta \phi(e,\mu)$ (bottom right). The notation of the of the lepton-pairs (bottom left) figure is the same as in Fig.~\ref{W3}.}
\label{W5}
\end{figure}

In Figs.~\ref{W4}~and~\ref{W5}, the differential cross-section for the production of $W^+W^-+jet(s)$ is plotted as functions of the transverse momenta of the leading leptons and the produced hadronic jets, the scalar sum of all jet transverse momenta as well as the invariant mass, the pseudorapidity and the azimuthal separation angle of the lepton-pairs. Similar to our initial impressions, the NLO collinear predictions from \textsf{Herwig~7} with both the \texttt{QCD$\oplus$QED} and the \texttt{QCD$\oplus$QED$\oplus$EW} parton shower choices are in agreement with the experimental measurements, while the inclusion of real EW radiations reduces the deviations of these theoretical predictions from ATLAS data. 

Several observations are made from the comparisons of Figs.~\ref{W4}~and~\ref{W5}:
(i) Although the ATLAS data universally favours the \texttt{Hw7.2 NLO $\otimes$ QCD$\oplus$QED$\oplus$EW} predictions, this agreement is more pronounced when the transverse momenta of the produced jets are large. This is since the negative EW virtual corrections become large in the high-$p_\top$ tails of these distributions and so the effect of positive real EW radiations becomes more visible, e.g. in the mass distributions of the produced lepton-pairs for $p_\top^{j} > 200$~GeV. 
(ii) Predictions from the $k_t$-factorisation framework also have better agreements with the experimental data in the high-$p_\top$ region. This can be contributed to the fact that AO constraint, which is a direct consequence of the colour-coherence effects of successive gluonic emissions, defines the characteristics of the KMR UPDFs and dramatically improves their high-energy behaviour~\cite{Modarres:2016hpe,Modarres:2016tow,Darvishi:2016fwo,Modarres:2018dwj,Darvishi:2019uzp}. 
(iii) In both figures, the KMR curves over-shoot the data in the low-$p_\top$ region with discrepancies up to $\sim$\%22 ($\sim$\%28) for the NLO (LO) predictions. This is most visible in the pseudorapidity and azimuthal angle separation distributions of Fig.~\ref{W5}. Such large differences may be the effect of asserting the AO cut off over the phase-space of the quark UPDFs in $qq'/qg \to W^\pm+jet(s)$ sub-processes that are dominant in the large-$x$ kinematic region. In fact, the misalignment of the KMR predictions with the experimental observations in the low-$p_\top$ region is a recurring limitation of this framework that has been reported and studied in the literature, see e.g.~\cite{Modarres:2016phz,Golec-Biernat:2018hqo,Hautmann:2019biw,Guiot:2019vsm}. The above argument is supported by observing the behaviour of the KMR curves in the top right panel of Fig.~\ref{W5}, where the insertion of a $p_\top^{j} > 200$~GeV cut suppresses the $qg/qq' \to W^\pm+jet(s)$ sub-processes, causing the $k_t$-factorisation predictions to substantially dip below the observed signal.   

\begin{figure}[t]
\centering
\includegraphics[width=0.49\textwidth]{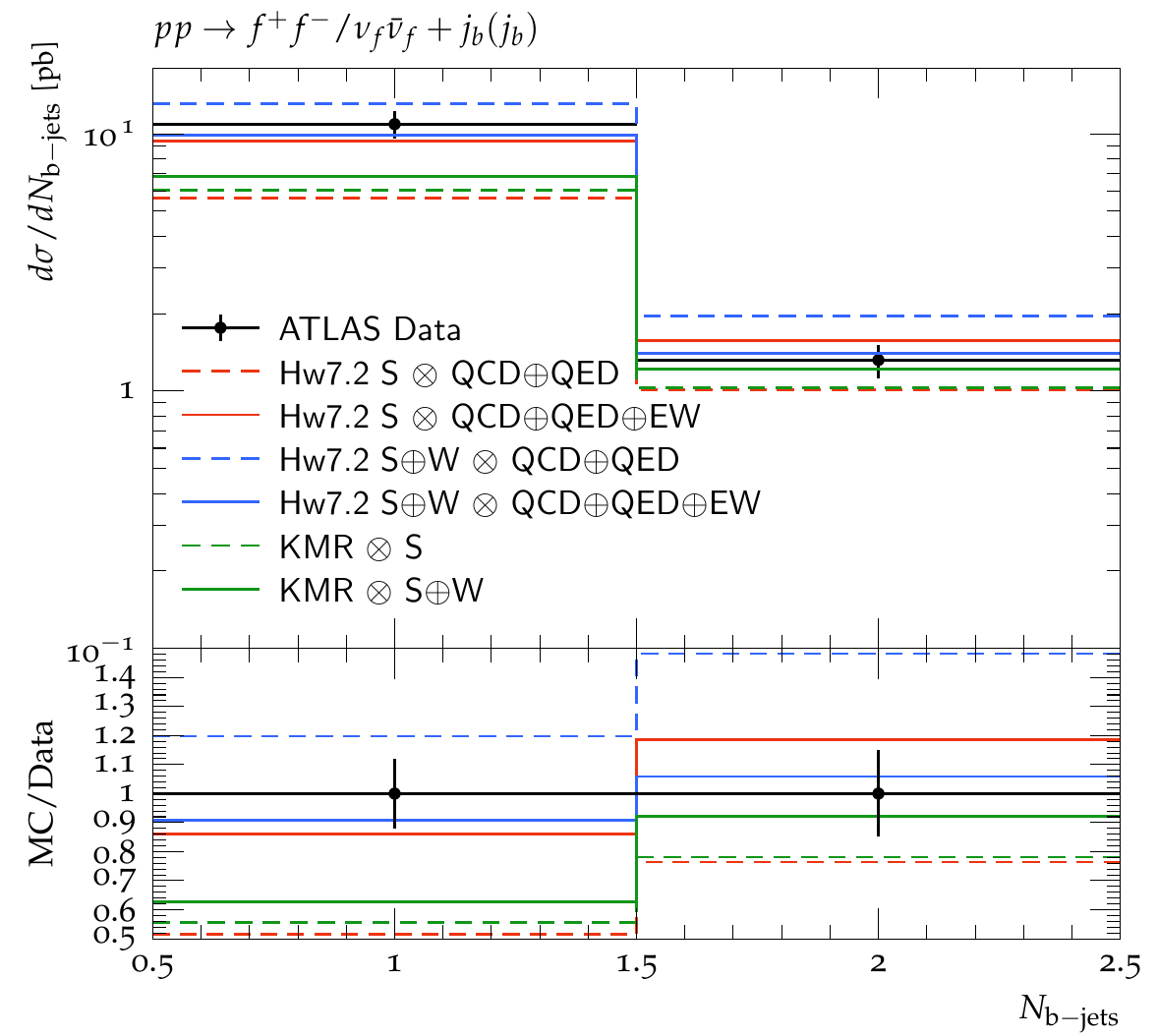}
\caption{\it Comparison of the differential $Z^0 + j_b (j_b)$ cross-section in various theoretical frameworks with ATLAS data~\cite{ATLAS:2020juj}. Differential cross-section of $Z^0 +j_b$ is plotted as functions of the pseudorapidity of the $Z^0$ boson (top left) and the leading $b$-tagged jet (top right). The distribution of $Z^0 +j_b j_b$ production rate is shown with respect to $\Delta y$ between two leading $b$-tagged jets (bottom left) and the invariant mass of the two leading $b$-tagged jet (bottom right). The notation of the lepton-pairs (bottom left) figure is the same as in Fig.~\ref{Z6}.}
\label{Z6}
\end{figure}

% Z+j production
The above comparisons clearly outline the importance of incorporating both QCD and EW real and virtual corrections on equal footing, while simulating EW-sensitive high-energy processes at the LHC. It would also be interesting to investigate the consequences of introducing real EW corrections, via EW partonic cascades, in the absence of virtual corrections. To this end, we look at the production of a single $Z^0$ boson alongside one or two $b$-tagged hadronic jets. Figs.~\ref{Z6} shows the differential $Z^0 + j_b (j_b)$ cross-section compared with ATLAS data~\cite{ATLAS:2020juj}. The theoretical predictions are generated using \textsf{Herwig~7}, with (marked with \texttt{S$\oplus$W}) or without (marked with \texttt{S}) virtual EW corrections in their NLO MEs, and enhanced with either \texttt{QCD$\oplus$QED} (dashed lines) or the \texttt{QCD$\oplus$QED$\oplus$EW} (solid lines) parton shower schemes. The $k_t$-factorisation predictions are presented with green histograms. These are generated with the use of purely loop-induced production channels up to two explicit $b$-jets, with (solid lines) or without (dashed lines) the inclusion of virtual EW corrections. 

\begin{figure}[t]
\centering
\includegraphics[width=1\textwidth]{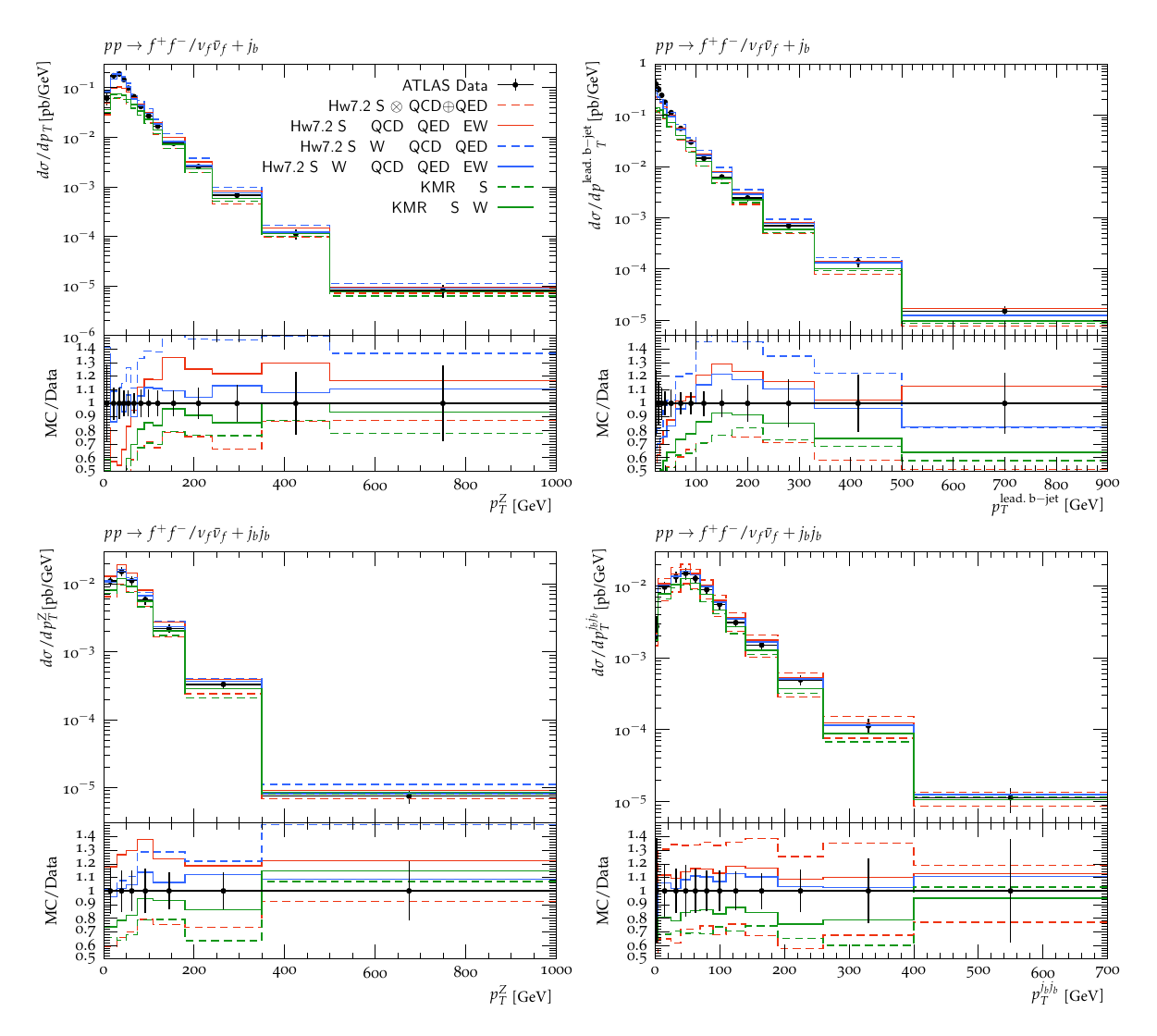}
\caption{\it Comparison of the differential $Z^0 + j_b (j_b)$ cross-section in various theoretical frameworks with ATLAS data~\cite{ATLAS:2020juj}. Differential cross-section of $Z^0 +j_b$ (top panels) and $Z^0 +j_b j_b$ (bottom panels) are plotted as functions of the transverse momenta of the $Z^0$ boson (left) and the $b$-tagged jet(s) (right). The notation of the of the lepton-pairs (bottom left) figure is the same as in Fig.~\ref{Z6}.}
\label{Z7}
\end{figure}

% observations 
From these comparisons, it can be readily seen that the \texttt{Hw7.2 S$\oplus$W $\otimes$ QCD$\oplus$QED$\oplus$EW} predictions are within the statistical+systematic uncertainty bounds of the data for both $Z^0 + j_b$ and $Z^0 + j_b j_b$ production events. Similar results in the absence of virtual EW corrections, \texttt{Hw7.2 S $\otimes$ QCD$\oplus$QED$\oplus$EW}, deviate from these uncertainty bounds with very small margins while the \texttt{Hw7.2 S$\oplus$W $\otimes$ QCD$\oplus$QED} predictions, in the absence of real EW corrections, overestimate the experimental measurements by $\sim + \%9$ for single and $\sim + \%19$ for double $b$-jet events. It is interesting to note that the use of \texttt{QCD$\oplus$QED$\oplus$EW} parton shower on \texttt{S$\oplus$W} MEs has reduced the predictions for the single and double $b$-jet events by $\sim - \%14$ and $\sim - \%17$, respectively. This suggests that in some regions of the phase-space, the sub-leading EW emissions were favoured over the QCD radiations, resulting in a reduction in the produced signals. 

\begin{figure}[t]
\centering
\includegraphics[width=1\textwidth]{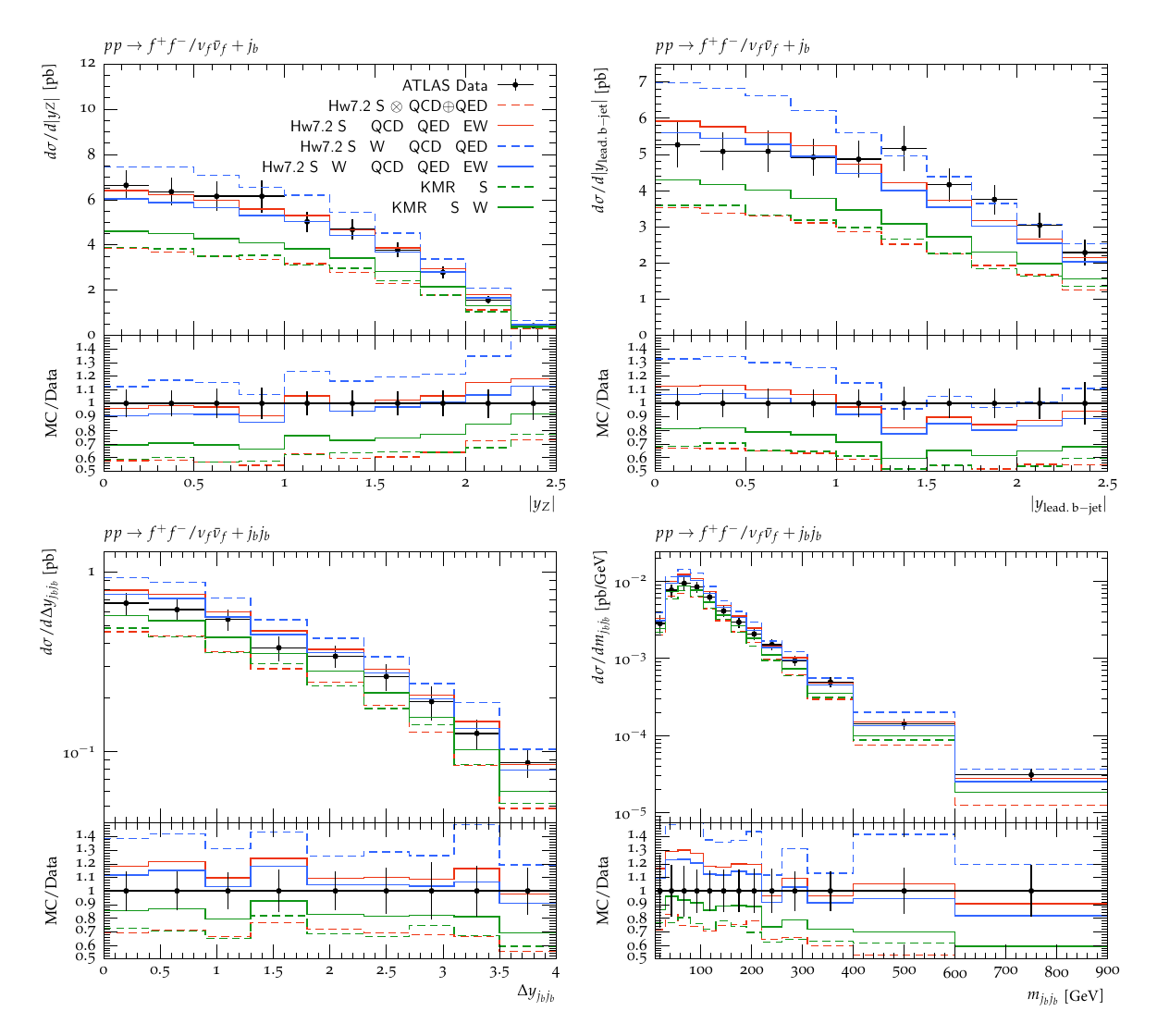}
\caption{\it Comparison of the differential $Z^0 + j_b (j_b)$ cross-section in various theoretical frameworks with ATLAS data~\cite{ATLAS:2020juj}. Differential cross-section of $Z^0 +j_b$ is plotted as functions of the pseudorapidity of the $Z^0$ boson (top left) and the leading $b$-tagged jet (top right). The distribution of $Z^0 +j_b j_b$ production rate is shown with respect to $\Delta y$ between two leading $b$-tagged jets (bottom left) and the invariant mass of the two leading $b$-tagged jet (bottom right). The notation of the lepton-pairs (bottom left) figure is the same as in Fig.~\ref{Z6}.}
\label{Z8}
\end{figure}

%7 8 figues
The above observations are complemented by the comparisons presented in Figs.~\ref{Z7} and \ref{Z8}, where the differential $Z^0 + j_b (j_b)$ cross-sections are plotted as functions of various physical observables, namely the transverse momenta and the pseudorapidities of the $Z^0$ boson and the $b$-tagged jet(s), $\Delta y$ between two leading $b$-tagged jets and their invariant mass in $Z^0 + j_b j_b$ events. In all of these results, the ATLAS data favours the simultaneous inclusion of both real and virtual EW corrections. It is also noticeable that the \texttt{Hw7.2 S $\otimes$ QCD$\oplus$QED$\oplus$EW} predictions, generated with only real EW emissions, present better descriptions of the experimental data, compared to the \texttt{Hw7.2 S$\oplus$W $\otimes$ QCD$\oplus$QED} results. This suggests that the inclusion of real EW emissions can take precedence over the incorporation of virtual EW corrections in the processes MEs, for the simulation of EW-sensitive production events.

One may note that in the above-mentioned predictions, the \texttt{KMR $\otimes$ S$\oplus$W} (\texttt{KMR $\otimes$ S}) curves underestimate the data throughout the phase-space by $\sim - \%19$ ($\sim - \%26$) for $Z^0+j_b$ and $\sim - \%7$ ($\sim - \%16$) for $Z^0+j_bj_b$ productions. This discrepancy between the KMR predictions and the measurements is again larger in the low-$p_\top$ region which is consistent with our previous observations on the limitations of the $k_t$-factorisation framework. Additionally, it can be seen that the inclusion of the EW virtual corrections has a smaller impact on the $k_t$-factorisation results compared to that of the collinear framework. This is since a large number of these one-loop EW corrections can be realized only with non-ladder-type Feynman diagrams that cannot accompany the $k_t$-factorisation UPDFs in the corresponding calculations and further enhancement is possible only when higher-order EW corrections are introduced into the MEs of the hard processes. Nevertheless, despite these shortcomings, the $k_t$-factorisation predictions in their rather minimalistic simulations show a reasonably good agreement with experimental data considering that they are produced without any parton shower enhancements.

\begin{figure}[t]
\centering
\includegraphics[width=1\textwidth]{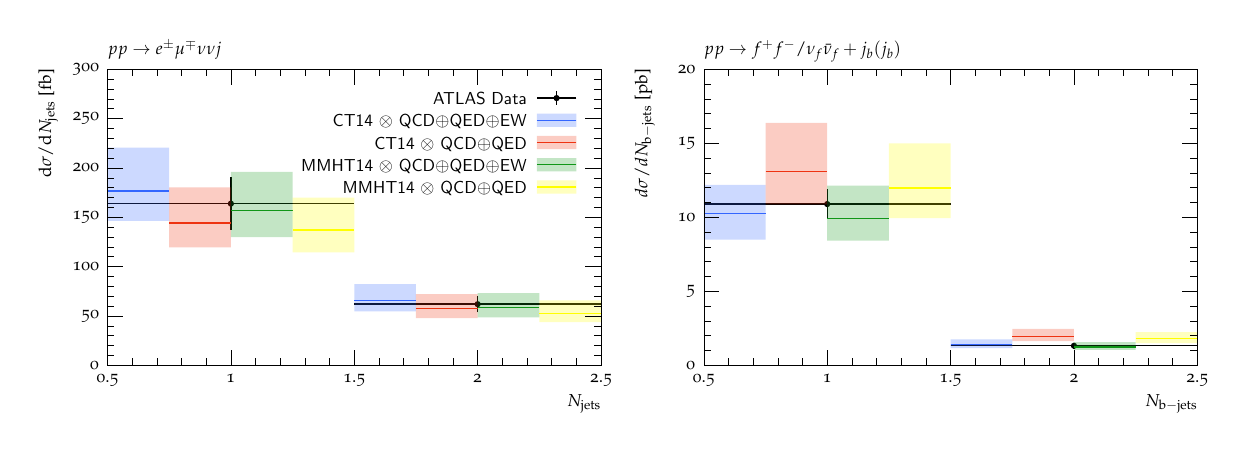}
\caption{\it Differential cross-sections for $W^+W^-+jet(s)$ (left panel) and $Z^0 + j_b (j_b)$ (right panel) productions as functions of the number of produced jets. Theoretical predictions are produced with NLO QCD+EW MEs, using \texttt{CT14} (blue and red histograms) and \texttt{MMHT14}~\cite{Harland-Lang:2014zoa} (green and yellow histograms) PDFs. The events are showered with either \texttt{QCD$\oplus$QED} (red and yellow histograms) or \texttt{QCD$\oplus$QED$\oplus$EW} (blue and green histograms) parton shower schemes. The uncertainty bounds are generated by manipulating renormalization, factorization and shower scales by a factor of $2$~\cite{Bellm:2016rhh}.}
\label{ScaleVary}
\end{figure}

Finally, in Fig.~\ref{ScaleVary} we study the effects of PDF selection and scale variation on the implemented parton shower schemes. To this end, we compare our previous results with NLO QCD+EW MEs and \texttt{CT14} PDFs against similar predictions with \texttt{MMHT14} PDFs~\cite{Harland-Lang:2014zoa}. We also manipulate the factorization (same as the renormalization) and the shower scales by a factor of $2$~\cite{Bellm:2016rhh} to generate the corresponding uncertainty bounds. It is observed that using different sets of PDFs could affect the predicted cross-sections by a non-negligible factor, as they determine the behaviour of the hard processes. On the other hand, the effects of the EW parton shower are unchanged. This is since the majority of the EW real corrections originate from the final-state radiations where the forward evolutions of the successive splittings are independent of the choice of the initial PDFs.

%final comments
In the above analysis, we have showcased the capability of the \texttt{QCD$\oplus$QED$\oplus$EW} parton shower scheme in \textsf{Herwig~7} general-purpose Monte-Carlo event generator and its importance in predicting the high-energy precision measurement data from the LHC. A particularly interesting observation in the present work was that treating EW real radiations with the same level as the QCD emissions can become more important than the inclusion of virtual EW corrections, especially in the EW-sensitive regions of the phase space.

\section{Summary and Conclusions}\label{Conc}

In this paper, we calculated the fiducial and differential cross-sections of $W^+W^-+jet(s)$ and $Z^0+jet(s)$ production events through $W^+ W^- \to e^+ \nu_e \; \mu^- \bar{\nu}_\mu$ and $Z^0 \to \ell^+ \ell^-/\nu_\ell \bar{\nu}_\ell$ decay modes, in the presence of EW corrections through hard EW boson self-interactions and/or real EW partonic cascade emissions. The calculations were carried out for proton–proton collisions at $\sqrt{s} = 13$~TeV, using \textsf{Herwig~7} general-purpose Monte-Carlo event generator with LO or NLO matrix elements and different parton-shower configurations, i.e. \texttt{QCD$\oplus$QED} and \texttt{QCD$\oplus$QED$\oplus$EW} schemes. The results were compared against experimental data from ATLAS and with computations from the $k_t$-factorisation framework using AO KMR UPDFs.

It was shown that the ATLAS precision measurements of the EW-sensitive $W^{\pm}/Z^0+jet(s)$ events universally favour the collinear predictions of \textsf{Herwig~7} that were enhanced with the simultaneous inclusion of real and virtual EW corrections. A particularly interesting observation was that treating EW real radiations with the same level as the QCD emissions can become more important than the inclusion of virtual EW corrections, especially in the EW-sensitive regions of the phase spaces. 

Additionally, we have demonstrated that the KMR $k_t$-factorisation framework, to a lesser degree, can produce reasonably good predictions compared to the experimental measurements without any additional parton shower enhancements. We have also argued that the limitations of the $k_t$-factorisation framework may be improved upon by introducing higher-order corrections to the hard MEs or by employing alternative transverse momentum PDFs that refine the low-$p_\top$ behaviour of the incoming partons in the $x \to 1$ kinematic region, e.g. the Parton Branching formalism~\cite{Hautmann:2017xtx, Hautmann:2017fcj, BermudezMartinez:2019anj, Martinez:2021chk}.

Finally, the capability of the \texttt{QCD$\oplus$QED$\oplus$EW} parton shower scheme in \textsf{Herwig~7} general-purpose Monte-Carlo event generator and its importance in predicting the high-energy precision measurement data from the LHC was outlined. This has turned out to be a crucial development that marks the next step in the evolution of high-energy Monte-Carlo simulators and can open the door to further parton-shower-related developments, e.g. beyond the SM particle cascades and Higgs boson self-interactions that will become relevant with the upcoming and inevitable push in the probe energies of the existing and future particle colliders. We intend to expand our investigation of the applications and the consequences of introducing EW parton shower enhancements through systematic comparisons between different algorithms, e,g, between AO and dipole EW parton showers in our future studies.

\section*{Acknowledgements}
\noindent
The authors sincerely thank Simon Pl\"atzer for instructive discussions on this work. \\
The work of ND is supported by the National Natural Science Foundation of China (NSFC) under grants No. 12022514, No. 11875003 and No. 12047503, and CAS Project for Young
Scientists in Basic Research YSBR-006, by the Development Program of China under Grant No. 2020YFC2201501 (2021/12/28) and by the CAS President’s International Fellowship Initiative (PIFI) grant. ND is also supported by the Polish National Science Centre HARMONIA grant under contract UMO-2015/20/M/ST2/00518 (2016-2021).
\\
MRM is supported by the UK Science and Technology Facilities Council (grant numbers ST/P001246/1). The work of MRM has also received funding from the European Union’s Horizon 2020 research and innovation program as part of the Marie Skłodowska-Curie Innovative Training Network MCnetITN3 (grant agreement no. 722104).


\begin{thebibliography}{99}
\addcontentsline{toc}{chapter}{Bibliographie}

\vspace{-0.2in}

%\cite{Denner:2019vbn}
\bibitem{Denner:2019vbn}
A.~Denner and S.~Dittmaier,
%``Electroweak Radiative Corrections for Collider Physics,''
Phys. Rept. \textbf{864} (2020), 1-163.
%doi:10.1016/j.physrep.2020.04.001
%[arXiv:1912.06823 [hep-ph]].
%65 citations counted in INSPIRE as of 22 Dec 2021

%\cite{Bellm:2017ktr}
\bibitem{Bellm:2017ktr}
J.~Bellm, S.~Gieseke and S.~Pl\"atzer,
%``Merging NLO Multi-jet Calculations with Improved Unitarization,''
Eur. Phys. J. C \textbf{78} (2018) no.3, 244.
%doi:10.1140/epjc/s10052-018-5723-2
%[arXiv:1705.06700 [hep-ph]].
%39 citations counted in INSPIRE as of 24 May 2022

%\cite{Gutschow:2018tuk}
\bibitem{Gutschow:2018tuk}
C.~G\"utschow, J.~M.~Lindert and M.~Sch\"onherr,
%``Multi-jet merged top-pair production including electroweak corrections,''
Eur. Phys. J. C \textbf{78} (2018) no.4, 317.
%doi:10.1140/epjc/s10052-018-5804-2
%[arXiv:1803.00950 [hep-ph]].
%38 citations counted in INSPIRE as of 24 May 2022

%\cite{Brauer:2020kfv}
\bibitem{Brauer:2020kfv}
S.~Br\"auer, A.~Denner, M.~Pellen, M.~Sch\"onherr and S.~Schumann,
%``Fixed-order and merged parton-shower predictions for WW and WWj production at the LHC including NLO QCD and EW corrections,''
JHEP \textbf{10} (2020), 159.
%doi:10.1007/JHEP10(2020)159
%[arXiv:2005.12128 [hep-ph]].
%16 citations counted in INSPIRE as of 24 May 2022

%\cite{Bothmann:2021led}
\bibitem{Bothmann:2021led}
E.~Bothmann, D.~Napoletano, M.~Sch\"onherr, S.~Schumann and S.~L.~Villani,
%``Higher-order EW corrections in $ZZ$ and $ZZj$ production at the LHC,''
[arXiv:2111.13453 [hep-ph]].
%0 citations counted in INSPIRE as of 27 Dec 2021

%\cite{Bahr:2008pv}
\bibitem{Bahr:2008pv} 
  M.~Bahr {\it et al.},
  %``Herwig++ Physics and Manual,''
  Eur.\ Phys.\ J.\ C {\bf 58}, 639 (2008).
  %doi:10.1140/epjc/s10052-008-0798-9
  %[arXiv:0803.0883 [hep-ph]].
  %%CITATION = doi:10.1140/epjc/s10052-008-0798-9;%%
  %1988 citations counted in INSPIRE as of 22 Oct 2019
  
%\cite{Bellm:2015jjp}
\bibitem{Bellm:2015jjp} 
  J.~Bellm {\it et al.},
  %``Herwig 7.0/Herwig++ 3.0 release note,''
  Eur.\ Phys.\ J.\ C {\bf 76}, no. 4, 196 (2016).
  %doi:10.1140/epjc/s10052-016-4018-8
  %[arXiv:1512.01178 [hep-ph]].
  %%CITATION = doi:10.1140/epjc/s10052-016-4018-8;%%
  %381 citations counted in INSPIRE as of 16 Oct 2019
  
%\cite{Bellm:2017bvx}
\bibitem{Bellm:2017bvx} 
  J.~Bellm {\it et al.},
  %``Herwig 7.1 Release Note,''
  arXiv:1705.06919 [hep-ph].
  %%CITATION = ARXIV:1705.06919;%%
  %52 citations counted in INSPIRE as of 16 Oct 2019

%\cite{Cullen:2014yla}
\bibitem{Cullen:2014yla}
G.~Cullen, H.~van Deurzen, N.~Greiner, G.~Heinrich, G.~Luisoni, P.~Mastrolia, E.~Mirabella, G.~Ossola, T.~Peraro and J.~Schlenk, \textit{et al.}
%``G$\scriptsize{O}$S$\scriptsize{AM}$-2.0: a tool for automated one-loop calculations within the Standard Model and beyond,''
Eur. Phys. J. C \textbf{74} (2014) no.8, 3001.
%doi:10.1140/epjc/s10052-014-3001-5
%[arXiv:1404.7096 [hep-ph]].
%255 citations counted in INSPIRE as of 23 Dec 2021

%\cite{Alwall:2014hca}
\bibitem{Alwall:2014hca}
J.~Alwall, R.~Frederix, S.~Frixione, V.~Hirschi, F.~Maltoni, O.~Mattelaer, H.~S.~Shao, T.~Stelzer, P.~Torrielli and M.~Zaro,
%``The automated computation of tree-level and next-to-leading order differential cross sections, and their matching to parton shower simulations,''
JHEP \textbf{07} (2014), 079.
%doi:10.1007/JHEP07(2014)079
%[arXiv:1405.0301 [hep-ph]].
%6200 citations counted in INSPIRE as of 23 Dec 2021

%\cite{Hirschi:2011pa}
\bibitem{Hirschi:2011pa}
V.~Hirschi, R.~Frederix, S.~Frixione, M.~V.~Garzelli, F.~Maltoni and R.~Pittau,
%``Automation of one-loop QCD corrections,''
JHEP \textbf{05} (2011), 044.
%doi:10.1007/JHEP05(2011)044
%[arXiv:1103.0621 [hep-ph]].
%458 citations counted in INSPIRE as of 23 Dec 2021

%\cite{Frederix:2018nkq}
\bibitem{Frederix:2018nkq}
R.~Frederix, S.~Frixione, V.~Hirschi, D.~Pagani, H.~S.~Shao and M.~Zaro,
%``The automation of next-to-leading order electroweak calculations,''
JHEP \textbf{07} (2018), 185
[erratum: JHEP \textbf{11} (2021), 085].
%doi:10.1007/JHEP11(2021)085
%[arXiv:1804.10017 [hep-ph]].
%140 citations counted in INSPIRE as of 23 Dec 2021

%\cite{CMS:2019kqw}
\bibitem{CMS:2019kqw} 
  CMS Collaboration [CMS Collaboration],
  %``Measurements of differential Higgs boson production cross sections in the leptonic WW decay mode at $\sqrt{s} = 13~\mathrm{TeV}$,''
  CMS-PAS-HIG-19-002.
  %%CITATION = CMS-PAS-HIG-19-002;%%
%\cite{Aaboud:2018gay}

\bibitem{Aaboud:2018gay} 
  M.~Aaboud {\it et al.} [ATLAS Collaboration],
  %``Search for Higgs bosons produced via vector-boson fusion and decaying into bottom quark pairs in $\sqrt{s} = 13$ $\mathrm{TeV}$ $pp$ collisions with the ATLAS detector,''
  Phys.\ Rev.\ D {\bf 98}, no. 5, 052003 (2018).
%doi:10.1103/PhysRevD.98.052003
%  [arXiv:1807.08639 [hep-ex]].
  %%CITATION = doi:10.1103/PhysRevD.98.052003;%%
  %13 citations counted in INSPIRE as of 17 Oct 2019

%\cite{Buccioni:2019sur}
\bibitem{Buccioni:2019sur}
F.~Buccioni, J.~N.~Lang, J.~M.~Lindert, P.~Maierh\"ofer, S.~Pozzorini, H.~Zhang and M.~F.~Zoller,
%``OpenLoops 2,''
Eur. Phys. J. C \textbf{79} (2019) no.10, 866.
%doi:10.1140/epjc/s10052-019-7306-2
%[arXiv:1907.13071 [hep-ph]].
%160 citations counted in INSPIRE as of 23 Dec 2021

%\cite{Beenakker:2000kb}
\bibitem{Beenakker:2000kb}
W.~Beenakker and A.~Werthenbach,
%``New insights into the perturbative structure of electroweak Sudakov logarithms,''
Phys. Lett. B \textbf{489} (2000), 148-156.
%doi:10.1016/S0370-2693(00)00900-X
%[arXiv:hep-ph/0005316 [hep-ph]].
%63 citations counted in INSPIRE as of 24 May 2022

%\cite{Schonherr:2017qcj}
\bibitem{Schonherr:2017qcj}
M.~Sch\"onherr,
%``An automated subtraction of NLO EW infrared divergences,''
Eur. Phys. J. C \textbf{78} (2018) no.2, 119.
%doi:10.1140/epjc/s10052-018-5600-z
%[arXiv:1712.07975 [hep-ph]].
%49 citations counted in INSPIRE as of 24 May 2022

%\cite{Christiansen:2014kba}
\bibitem{Christiansen:2014kba}
J.~R.~Christiansen and T.~Sj\"ostrand,
%``Weak Gauge Boson Radiation in Parton Showers,''
JHEP \textbf{04} (2014), 115.
%doi:10.1007/JHEP04(2014)115
%[arXiv:1401.5238 [hep-ph]].
%71 citations counted in INSPIRE as of 24 May 2022

%\cite{Chiesa:2013yma}
\bibitem{Chiesa:2013yma} 
  M.~Chiesa, G.~Montagna, L.~Barzè, M.~Moretti, O.~Nicrosini, F.~Piccinini and F.~Tramontano,
  %``Electroweak Sudakov Corrections to New Physics Searches at the LHC,''
  Phys.\ Rev.\ Lett.\  {\bf 111}, no. 12, 121801 (2013).
  %doi:10.1103/PhysRevLett.111.121801
  %[arXiv:1305.6837 [hep-ph]].
  %%CITATION = doi:10.1103/PhysRevLett.111.121801;%%
  %34 citations counted in INSPIRE as of 17 Oct 2019
  
%\cite{Christiansen:2014kba}
\bibitem{Christiansen:2014kba} 
  J.~R.~Christiansen and T.~Sjöstrand,
  %``Weak Gauge Boson Radiation in Parton Showers,''
  JHEP {\bf 1404}, 115 (2014).
  %doi:10.1007/JHEP04(2014)115
  %[arXiv:1401.5238 [hep-ph]].
  %%CITATION = doi:10.1007/JHEP04(2014)115;%%
  %46 citations counted in INSPIRE as of 17 Oct 2019
  
%\cite{Krauss:2014yaa}
\bibitem{Krauss:2014yaa} 
  F.~Krauss, P.~Petrov, M.~Schoenherr and M.~Spannowsky,
  %``Measuring collinear W emissions inside jets,''
  Phys.\ Rev.\ D {\bf 89}, no. 11, 114006 (2014).
  %doi:10.1103/PhysRevD.89.114006
  %[arXiv:1403.4788 [hep-ph]].
  %%CITATION = doi:10.1103/PhysRevD.89.114006;%%
  %25 citations counted in INSPIRE as of 17 Oct 2019
  
%\cite{Chen:2016wkt}
\bibitem{Chen:2016wkt}
J.~Chen, T.~Han and B.~Tweedie,
%``Electroweak Splitting Functions and High Energy Showering,''
JHEP \textbf{11} (2017), 093.
%doi:10.1007/JHEP11(2017)093
%[arXiv:1611.00788 [hep-ph]].
%67 citations counted in INSPIRE as of 24 May 2022
  
%\cite{Mangano:2002ea}
\bibitem{Mangano:2002ea} 
  M.~L.~Mangano, M.~Moretti, F.~Piccinini, R.~Pittau and A.~D.~Polosa,
  %``ALPGEN, a generator for hard multiparton processes in hadronic collisions,''
  JHEP {\bf 0307}, 001 (2003).
  %doi:10.1088/1126-6708/2003/07/001
  %[hep-ph/0206293].
  %%CITATION = doi:10.1088/1126-6708/2003/07/001;%%
  %3255 citations counted in INSPIRE as of 17 Oct 2019
  
%\cite{Kleiss:2020rcg}
\bibitem{Kleiss:2020rcg} 
  R.~Kleiss and R.~Verheyen,
  %``Electroweak Radiation in Antenna Parton Showers,''
  arXiv:2002.09248 [hep-ph].
  %%CITATION = ARXIV:2002.09248;%%
  %1 citations counted in INSPIRE as of 03 Mar 2020
  
%\cite{Brooks:2021kji}
\bibitem{Brooks:2021kji}
H.~Brooks, P.~Skands and R.~Verheyen,
%``Interleaved Resonance Decays and Electroweak Radiation in Vincia,''
[arXiv:2108.10786 [hep-ph]].
%0 citations counted in INSPIRE as of 02 Sep 2021

%\cite{Pagani:2021vyk}
\bibitem{Pagani:2021vyk}
D.~Pagani and M.~Zaro,
%``One-loop electroweak Sudakov logarithms: a revisitation and automation,''
[arXiv:2110.03714 [hep-ph]].
%1 citations counted in INSPIRE as of 23 Dec 2021
  
%\cite{Masouminia:2021kne}
\bibitem{Masouminia:2021kne}
M.~R.~Masouminia and P.~Richardson,
%``Implementation of angularly ordered electroweak parton shower in Herwig 7,''
JHEP \textbf{04} (2022), 112.
%doi:10.1007/JHEP04(2022)112
%[arXiv:2108.10817 [hep-ph]].
%8 citations counted in INSPIRE as of 20 May 2022

%\cite{ATLAS:2020juj}
\bibitem{ATLAS:2020juj}
G.~Aad \textit{et al.} [ATLAS],
%``Measurements of the production cross-section for a $Z$ boson in association with $b$-jets in proton-proton collisions at $\sqrt{s} = 13$ TeV with the ATLAS detector,''
JHEP \textbf{07} (2020), 044.
%doi:10.1007/JHEP07(2020)044
%[arXiv:2003.11960 [hep-ex]].
%24 citations counted in INSPIRE as of 23 Dec 2021

%\cite{ATLAS:2021jgw}
\bibitem{ATLAS:2021jgw}
G.~Aad \textit{et al.} [ATLAS],
%``Measurements of $W^+W^-+\ge 1~$jet production cross-sections in $pp$ collisions at $\sqrt{s}=13~$TeV with the ATLAS detector,''
JHEP \textbf{06} (2021), 003.
%doi:10.1007/JHEP06(2021)003
%[arXiv:2103.10319 [hep-ex]].
%9 citations counted in INSPIRE as of 23 Dec 2021

%\cite{Catani:1990xk}
\bibitem{Catani:1990xk}
S.~Catani, M.~Ciafaloni and F.~Hautmann,
%``GLUON CONTRIBUTIONS TO SMALL x HEAVY FLAVOR PRODUCTION,''
Phys. Lett. B \textbf{242} (1990), 97-102.
%doi:10.1016/0370-2693(90)91601-7
%624 citations counted in INSPIRE as of 16 May 2022

%\cite{Catani:1990eg}
\bibitem{Catani:1990eg}
S.~Catani, M.~Ciafaloni and F.~Hautmann,
%``High-energy factorization and small x heavy flavor production,''
Nucl. Phys. B \textbf{366} (1991), 135-188.
%doi:10.1016/0550-3213(91)90055-3
%1191 citations counted in INSPIRE as of 16 May 2022

%\cite{Catani:1993ww}
\bibitem{Catani:1993ww}
S.~Catani, M.~Ciafaloni and F.~Hautmann,
%``High-energy factorization in QCD and minimal subtraction scheme,''
Phys. Lett. B \textbf{307} (1993), 147-153.
%doi:10.1016/0370-2693(93)90204-U
%162 citations counted in INSPIRE as of 16 May 2022

%\cite{Kimber:2001sc}
\bibitem{Kimber:2001sc}
M.~A.~Kimber, A.~D.~Martin and M.~G.~Ryskin,
%``Unintegrated parton distributions,''
Phys. Rev. D \textbf{63} (2001), 114027.
%doi:10.1103/PhysRevD.63.114027
%[arXiv:hep-ph/0101348 [hep-ph]].
%421 citations counted in INSPIRE as of 23 Dec 2021

%\cite{Martin:2009ii}
\bibitem{Martin:2009ii}
A.~D.~Martin, M.~G.~Ryskin and G.~Watt,
%``NLO prescription for unintegrated parton distributions,''
Eur. Phys. J. C \textbf{66} (2010), 163-172.
%doi:10.1140/epjc/s10052-010-1242-5
%[arXiv:0909.5529 [hep-ph]].
%112 citations counted in INSPIRE as of 23 Dec 2021

%\cite{Modarres:2016hpe}
\bibitem{Modarres:2016hpe}
M.~Modarres, M.~R.~Masouminia, R.~Aminzadeh Nik, H.~Hosseinkhani and N.~Olanj,
%``NLO production of $W^{\pm}$ and $Z^0$ vector bosons via hadron collisions in the frameworks of Kimber-Martin-Ryskin and Martin-Ryskin-Watt unintegrated parton distribution functions,''
Phys. Rev. D \textbf{94} (2016) no.7, 074035.
%doi:10.1103/PhysRevD.94.074035
%[arXiv:1609.07920 [hep-ph]].
%13 citations counted in INSPIRE as of 23 Dec 2021

%\cite{Modarres:2016tow}
\bibitem{Modarres:2016tow}
M.~Modarres, M.~R.~Masouminia, R.~Aminzadeh Nik, H.~Hosseinkhani and N.~Olanj,
%``$KMR$ $k_t$-factorization procedure for the description of the $LHCb$ forward hadron-hadron $Z^0$ production at $\sqrt{s}=13\;TeV$,''
Phys. Lett. B \textbf{772} (2017), 534-541.
%doi:10.1016/j.physletb.2017.07.015
%[arXiv:1610.02635 [hep-ph]].
%14 citations counted in INSPIRE as of 23 Dec 2021

%\cite{Darvishi:2016fwo}
\bibitem{Darvishi:2016fwo}
N.~Darvishi and M.~R.~Masouminia,
%``A phenomenological study on the production of Higgs bosons in the cSMCS model at the LHC,''
Nucl. Phys. B \textbf{923} (2017), 491-507.
%doi:10.1016/j.nuclphysb.2017.08.013
%[arXiv:1611.03312 [hep-ph]].
%5 citations counted in INSPIRE as of 23 Dec 2021

%\cite{Modarres:2018dwj}
\bibitem{Modarres:2018dwj}
M.~Modarres, M.~R.~Masouminia, R.~Aminzadeh Nik, H.~Hosseinkhani and N.~Olanj,
%``Semi- $NLO$ production of Higgs bosons in the framework of $k_t$ -factorization using $KMR$ unintegrated parton distributions,''
Nucl. Phys. B \textbf{926} (2018), 406-426.
%doi:10.1016/j.nuclphysb.2017.11.015
%10 citations counted in INSPIRE as of 23 Dec 2021

%\cite{Darvishi:2019uzp}
\bibitem{Darvishi:2019uzp}
N.~Darvishi, M.~R.~Masouminia and K.~Ostrolenk,
%``$k_t$ factorization versus collinear factorization in $W^{+}W^{-}$ pair production at the LHC,''
Phys. Rev. D \textbf{101} (2020) no.1, 014007.
%doi:10.1103/PhysRevD.101.014007
%[arXiv:1909.13862 [hep-ph]].
%5 citations counted in INSPIRE as of 23 Dec 2021

%\cite{Darvishi:2020paz}
\bibitem{Darvishi:2020paz}
N.~Darvishi and M.~R.~Masouminia,
%``Signature of the Maximally Symmetric 2HDM via $W^{\pm}/Z$-Quadruplet Productions at the LHC,''
Phys. Rev. D \textbf{103} (2021) no.9, 095031.
%doi:10.1103/PhysRevD.103.095031
%[arXiv:2012.14746 [hep-ph]].
%1 citations counted in INSPIRE as of 23 Dec 2021

%\cite{Baranov:2008hj}
\bibitem{Baranov:2008hj}
S.~P.~Baranov, A.~V.~Lipatov and N.~P.~Zotov,
%``Production of electroweak gauge bosons in off-shell gluon-gluon fusion,''
Phys. Rev. D \textbf{78} (2008), 014025.
%doi:10.1103/PhysRevD.78.014025
%[arXiv:0805.4821 [hep-ph]].
%37 citations counted in INSPIRE as of 25 Dec 2021

%\cite{Dulat:2015mca}
\bibitem{Dulat:2015mca}
S.~Dulat, T.~J.~Hou, J.~Gao, M.~Guzzi, J.~Huston, P.~Nadolsky, J.~Pumplin, C.~Schmidt, D.~Stump and C.~P.~Yuan,
%``New parton distribution functions from a global analysis of quantum chromodynamics,''
Phys. Rev. D \textbf{93} (2016) no.3, 033006.
%doi:10.1103/PhysRevD.93.033006
%[arXiv:1506.07443 [hep-ph]].
%1497 citations counted in INSPIRE as of 25 Dec 2021

%\cite{Hou:2016sho}
\bibitem{Hou:2016sho}
T.~J.~Hou, J.~Gao, J.~Huston, P.~Nadolsky, C.~Schmidt, D.~Stump, B.~T.~Wang, K.~P.~Xie, S.~Dulat and J.~Pumplin, \textit{et al.}
%``Reconstruction of Monte Carlo replicas from Hessian parton distributions,''
JHEP \textbf{03} (2017), 099.
%doi:10.1007/JHEP03(2017)099
%[arXiv:1607.06066 [hep-ph]].
%32 citations counted in INSPIRE as of 25 Dec 2021

%\cite{Modarres:2016phz}
\bibitem{Modarres:2016phz}
M.~Modarres, M.~R.~Masouminia, R.~Aminzadeh~Nik, H.~Hosseinkhani and N.~Olanj,
%``LHC production of forward-center and forward-forward di-jets in the $k_t$-factorization and transverse dependent unintegrated parton distribution frameworks,''
Nucl. Phys. B \textbf{922} (2017), 94-112.
%[arXiv:1610.02777 [hep-ph]].
%10 citations counted in INSPIRE as of 20 May 2022

%\cite{Golec-Biernat:2018hqo}
\bibitem{Golec-Biernat:2018hqo}
K.~Golec-Biernat and A.~M.~Stasto,
%``On the use of the KMR unintegrated parton distribution functions,''
Phys. Lett. B \textbf{781} (2018), 633-638.
%doi:10.1016/j.physletb.2018.04.061
[%arXiv:1803.06246 [hep-ph]].
%23 citations counted in INSPIRE as of 16 May 2022

%\cite{Martin:2009ii}
\bibitem{Martin:2009ii}
A.~D.~Martin, M.~G.~Ryskin and G.~Watt,
%``NLO prescription for unintegrated parton distributions,''
Eur. Phys. J. C \textbf{66} (2010), 163-172.
%doi:10.1140/epjc/s10052-010-1242-5
%[arXiv:0909.5529 [hep-ph]].
%115 citations counted in INSPIRE as of 20 May 2022

%\cite{Darvishi:2022gqt}
\bibitem{Darvishi:2022gqt}
N.~Darvishi, S.~H\"oche, J.~I.~M.~R.~Masouminia, Z.~Nagy, P.~Richardson and D.~E.~Soper,
%``Future prospects for parton showers,''
[arXiv:2203.06799 [hep-ph]].
%0 citations counted in INSPIRE as of 25 May 2022

%\cite{Campbell:2022qmc}
\bibitem{Campbell:2022qmc}
J.~M.~Campbell, M.~Diefenthaler, T.~J.~Hobbs, S.~H\"oche, J.~Isaacson, F.~Kling, S.~Mrenna, J.~Reuter, S.~Alioli and J.~R.~Andersen, \textit{et al.}
%``Event Generators for High-Energy Physics Experiments,''
[arXiv:2203.11110 [hep-ph]].
%6 citations counted in INSPIRE as of 25 May 2022

%\cite{Frixione:2022ofv}
\bibitem{Frixione:2022ofv}
S.~Frixione, E.~Laenen, C.~M.~C.~Calame, A.~Denner, S.~Dittmaier, T.~Engel, L.~Flower, L.~Gellersen, S.~Hoeche and S.~Jadach, \textit{et al.}
%``Initial state QED radiation aspects for future $e^+e^-$ colliders,''
[arXiv:2203.12557 [hep-ph]].
%1 citations counted in INSPIRE as of 25 May 2022

%\cite{Feng:2022inv}
\bibitem{Feng:2022inv}
J.~L.~Feng, F.~Kling, M.~H.~Reno, J.~Rojo, D.~Soldin, L.~A.~Anchordoqui, J.~Boyd, A.~Ismail, L.~Harland-Lang and K.~J.~Kelly, \textit{et al.}
%``The Forward Physics Facility at the High-Luminosity LHC,''
[arXiv:2203.05090 [hep-ex]].
%25 citations counted in INSPIRE as of 25 May 2022

%\cite{Gauld:2020deh}
\bibitem{Gauld:2020deh}
R.~Gauld, A.~Gehrmann-De Ridder, E.~W.~N.~Glover, A.~Huss and I.~Majer,
%``Predictions for $Z$ -Boson Production in Association with a $b$-Jet at $\mathcal {O}(\alpha_s^3)$,''
Phys. Rev. Lett. \textbf{125} (2020) no.22, 222002.
%doi:10.1103/PhysRevLett.125.222002
%[arXiv:2005.03016 [hep-ph]].
%17 citations counted in INSPIRE as of 24 May 2022

%\cite{Baranov:2021uol}
\bibitem{Baranov:2021uol}
S.~Baranov, A.~Bermudez Martinez, L.~I.~Estevez Banos, F.~Guzman, F.~Hautmann, H.~Jung, A.~Lelek, J.~Lidrych, A.~Lipatov and M.~Malyshev, \textit{et al.}
%``CASCADE3 A Monte Carlo event generator based on TMDs,''
Eur. Phys. J. C \textbf{81} (2021) no.5, 425
doi:10.1140/epjc/s10052-021-09203-8
[arXiv:2101.10221 [hep-ph]].
%44 citations counted in INSPIRE as of 17 Nov 2022

%\cite{Jung:2010si}
\bibitem{Jung:2010si}
H.~Jung, S.~Baranov, M.~Deak, A.~Grebenyuk, F.~Hautmann, M.~Hentschinski, A.~Knutsson, M.~Kramer, K.~Kutak and A.~Lipatov, \textit{et al.}
%``The CCFM Monte Carlo generator CASCADE version 2.2.03,''
Eur. Phys. J. C \textbf{70} (2010), 1237-1249
doi:10.1140/epjc/s10052-010-1507-z
[arXiv:1008.0152 [hep-ph]].
%228 citations counted in INSPIRE as of 17 Nov 2022

%\cite{BermudezMartinez:2022bpj}
\bibitem{BermudezMartinez:2022bpj}
A.~Bermudez Martinez, F.~Hautmann and M.~L.~Mangano,
%``Multi-jet merging with TMD parton branching,''
JHEP \textbf{09} (2022), 060
doi:10.1007/JHEP09(2022)060
[arXiv:2208.02276 [hep-ph]].
%6 citations counted in INSPIRE as of 17 Nov 2022

%\cite{Yang:2022qgk}
\bibitem{Yang:2022qgk}
H.~Yang, A.~Bermudez Martinez, L.~I.~E.~Banos, F.~Hautmann, H.~Jung, M.~Mendizabal, K.~M.~Figueroa, S.~Prestel, S.~T.~Monfared and A.~M.~van Kampen, \textit{et al.}
%``Back-to-back azimuthal correlations in $\mathrm {Z} +$jet events at high transverse momentum in the TMD parton branching method at next-to-leading order,''
Eur. Phys. J. C \textbf{82} (2022) no.8, 755
doi:10.1140/epjc/s10052-022-10715-0
[arXiv:2204.01528 [hep-ph]].
%8 citations counted in INSPIRE as of 17 Nov 2022

%\cite{Hautmann:2008vd}
\bibitem{Hautmann:2008vd}
F.~Hautmann and H.~Jung,
%``Angular correlations in multi-jet final states from k-perpendicular - dependent parton showers,''
JHEP \textbf{10} (2008), 113
doi:10.1088/1126-6708/2008/10/113
[arXiv:0805.1049 [hep-ph]].
%97 citations counted in INSPIRE as of 17 Nov 2022

%\cite{Hautmann:2019biw}
\bibitem{Hautmann:2019biw}
F.~Hautmann, L.~Keersmaekers, A.~Lelek and A.~M.~Van Kampen,
%``Dynamical resolution scale in transverse momentum distributions at the LHC,''
Nucl. Phys. B \textbf{949} (2019), 114795.
%doi:10.1016/j.nuclphysb.2019.114795
%[arXiv:1908.08524 [hep-ph]].
%30 citations counted in INSPIRE as of 16 May 2022

%\cite{Guiot:2019vsm}
\bibitem{Guiot:2019vsm}
B.~Guiot,
%``Pathologies of the Kimber-Martin-Ryskin prescriptions for unintegrated PDFs: Which prescription should be preferred?,''
Phys. Rev. D \textbf{101} (2020) no.5, 054006.
%doi:10.1103/PhysRevD.101.054006
%[arXiv:1910.09656 [hep-ph]].
%10 citations counted in INSPIRE as of 16 May 2022

%\cite{Hautmann:2017xtx}
\bibitem{Hautmann:2017xtx}
F.~Hautmann, H.~Jung, A.~Lelek, V.~Radescu and R.~Zlebcik,
%``Soft-gluon resolution scale in QCD evolution equations,''
Phys. Lett. B \textbf{772} (2017), 446-451.
%doi:10.1016/j.physletb.2017.07.005
%[arXiv:1704.01757 [hep-ph]].
%85 citations counted in INSPIRE as of 20 May 2022

%\cite{Hautmann:2017fcj}
\bibitem{Hautmann:2017fcj}
F.~Hautmann, H.~Jung, A.~Lelek, V.~Radescu and R.~Zlebcik,
%``Collinear and TMD Quark and Gluon Densities from Parton Branching Solution of QCD Evolution Equations,''
JHEP \textbf{01} (2018), 070.
%doi:10.1007/JHEP01(2018)070
%[arXiv:1708.03279 [hep-ph]].
%96 citations counted in INSPIRE as of 20 May 2022

%\cite{BermudezMartinez:2019anj}
\bibitem{BermudezMartinez:2019anj}
A.~Bermudez Martinez, P.~Connor, D.~Dominguez Damiani, L.~I.~Estevez Banos, F.~Hautmann, H.~Jung, J.~Lidrych, M.~Schmitz, S.~Taheri Monfared and Q.~Wang, \textit{et al.}
%``Production of Z-bosons in the parton branching method,''
Phys. Rev. D \textbf{100} (2019) no.7, 074027.
%doi:10.1103/PhysRevD.100.074027
%[arXiv:1906.00919 [hep-ph]].
%52 citations counted in INSPIRE as of 21 May 2022

%\cite{Martinez:2021chk}
\bibitem{Martinez:2021chk}
A.~B.~Martinez, F.~Hautmann and M.~L.~Mangano,
%``TMD evolution and multi-jet merging,''
Phys. Lett. B \textbf{822} (2021), 136700.
%doi:10.1016/j.physletb.2021.136700
%[arXiv:2107.01224 [hep-ph]].
%14 citations counted in INSPIRE as of 21 May 2022

%\cite{Hautmann:2014kza}
\bibitem{Hautmann:2014kza}
F.~Hautmann, H.~Jung, M.~Kr\"amer, P.~J.~Mulders, E.~R.~Nocera, T.~C.~Rogers and A.~Signori,
%``TMDlib and TMDplotter: library and plotting tools for transverse-momentum-dependent parton distributions,''
Eur. Phys. J. C \textbf{74} (2014), 3220.
%doi:10.1140/epjc/s10052-014-3220-9
%[arXiv:1408.3015 [hep-ph]].
%101 citations counted in INSPIRE as of 16 May 2022

%\cite{Abdulov:2021ivr}
\bibitem{Abdulov:2021ivr}
N.~A.~Abdulov, A.~Bacchetta, S.~Baranov, A.~B.~Martinez, V.~Bertone, C.~Bissolotti, V.~Candelise, L.~I.~E.~Banos, M.~Bury and P.~L.~S.~Connor, \textit{et al.}
%``TMDlib2 and TMDplotter: a platform for 3D hadron structure studies,''
Eur. Phys. J. C \textbf{81} (2021) no.8, 752.
%doi:10.1140/epjc/s10052-021-09508-8
%[arXiv:2103.09741 [hep-ph]].
%23 citations counted in INSPIRE as of 16 May 2022

%\cite{Platzer:2011bc}
\bibitem{Platzer:2011bc}
S.~Platzer and S.~Gieseke,
%``Dipole Showers and Automated NLO Matching in Herwig++,''
Eur. Phys. J. C \textbf{72} (2012), 2187
%doi:10.1140/epjc/s10052-012-2187-7
%[arXiv:1109.6256 [hep-ph]].
%167 citations counted in INSPIRE as of 24 May 2022

%\cite{Buckley:2010ar}
\bibitem{Buckley:2010ar}
A.~Buckley, J.~Butterworth, D.~Grellscheid, H.~Hoeth, L.~Lonnblad, J.~Monk, H.~Schulz and F.~Siegert,
%``Rivet user manual,''
Comput. Phys. Commun. \textbf{184} (2013), 2803-2819.
%doi:10.1016/j.cpc.2013.05.021
%[arXiv:1003.0694 [hep-ph]].
%685 citations counted in INSPIRE as of 26 Dec 2021

\bibitem{FORM} J.A.M. Vermaseren, 
%Symbolic Manipulation with FORM, published by 
Computer Algebra, 
Nederland, Kruislaan 413, 1098, SJ Amsterdaam, 
1991; ISBN 90-74116-01-9. 

%\cite{Harland-Lang:2014zoa}
\bibitem{Harland-Lang:2014zoa}
L.~A.~Harland-Lang, A.~D.~Martin, P.~Motylinski and R.~S.~Thorne,
%``Parton distributions in the LHC era: MMHT 2014 PDFs,''
Eur. Phys. J. C \textbf{75}, no.5, 204 (2015).
%doi:10.1140/epjc/s10052-015-3397-6
%[arXiv:1412.3989 [hep-ph]].
%1527 citations counted in INSPIRE as of 23 Jul 2022

%\cite{Bellm:2016rhh}
\bibitem{Bellm:2016rhh}
J.~Bellm, G.~Nail, S.~Pl\"atzer, P.~Schichtel and A.~Si\'odmok,
%``Parton Shower Uncertainties with Herwig 7: Benchmarks at Leading Order,''
Eur. Phys. J. C \textbf{76}, no.12, 665 (2016).
%doi:10.1140/epjc/s10052-016-4506-x
%[arXiv:1605.01338 [hep-ph]].
%34 citations counted in INSPIRE as of 23 Jul 2022

\end{thebibliography}
\end{document}